\documentclass{article}

\usepackage{amssymb}
\usepackage{amsmath}
\usepackage{amsthm}
\usepackage{stmaryrd}
\usepackage[usenames,dvipsnames]{xcolor}
\usepackage{enumerate}
\usepackage{proof-dashed}
\usepackage{todonotes} 
\usepackage{listings}
\usepackage{import}
\usepackage{wrapfig}
\usepackage{breakcites}
\usepackage{hyperref}
\usepackage[all]{xy}
\usepackage{booktabs}
\usepackage{comment}   
\usepackage{authblk}

\usepackage{pgfplots}
\pgfplotsset{compat=1.3}

\newtheorem{theorem}{Theorem}
\newtheorem{definition}{Definition}
\newtheorem{lemma}{Lemma}
\newtheorem{example}{Example}

\newtheorem{remark}{Remark}

\newcommand{\kim}[1]{\todo[color=yellow]{\textbf{K:} {#1}}}

\newcommand{\m}[1]{\ensuremath{\mathsf{#1}}}
\newcommand{\code}[1]{\texttt{\upshape #1}}
\newcommand{\tuple}[1]{\ensuremath{\langle{#1}\rangle}}
\newcommand{\til}{\tilde}
\newcommand{\eoe}{\hfill$\triangleleft$}
\newcommand{\best}[1]{\textcolor{mygreen}{\textbf{#1}}}
\newcommand{\worst}[1]{\textcolor{myred}{\textbf{#1}}}
\definecolor{mygreen}{rgb}{0,.7,0}
\definecolor{myred}{rgb}{.7,0,0}

\newcommand{\pid}[1]{\m{#1}}
\newcommand{\nil}{\boldsymbol 0}
\newcommand{\procs}[1][]{\mathcal D_{\pid{#1}}}
\newcommand{\lto}[2][\procs]{\xrightarrow{#2}_{#1}}
\newcommand{\abscom}[2]{{#1}\;\code{-\hspace{-0.3mm}>}\;{#2}}
\newcommand{\lcom}[3]{\abscom{\pid{#1}.{#2}}{\pid{#3}}}
\newcommand{\lmto}[1]{\xrightarrow{\smash{#1}}^\ast}
\newcommand{\lcmto}[2][\procs]{\xrightarrow{\smash{#2}}^\ast_{#1}}
\newcommand{\lleft}{\textsc{l}}
\newcommand{\lright}{\textsc{r}}

\newcommand{\bsend}[2]{{\pid #1}!\mathit{#2}}
\newcommand{\brecv}[2]{\pid #1?{\mathit{#2}}}
\newcommand{\procterm}[2]{\m{def} \, #1 \, \m{in} \, #2}
\newcommand{\bsel}[2]{{\pid #1}\oplus\mathit{#2}}
\newcommand{\bbranch}[2]{{\pid #1}\&\{{#2}\}}
\newcommand{\bcond}[3]{\m{if}\, \mathit{#1} \, \m{then} \, #2 \, \m{else} \, #3}
\newcommand{\brec}[2]{#1  =  #2}
\newcommand{\actor}[3][]{\pid{#2} \mathrel{\triangleright_{#1}} #3}
\newcommand{\parp}{\, \boldsymbol{|} \, }

\newcommand{\com}[4]{\abscom{{\pid #1}.\mathit{#2}}{\pid{#3}.\mathit{#4}}}
\newcommand{\gencom}{\com peqx}
\newcommand{\sel}[3]{\abscom{\pid{#1}}{\pid{#2}[\mathit{#3}]}}
\newcommand{\gensel}{\sel{\pid p}{\pid q}{\ell}}
\newcommand{\cond}[4]{\m{if}\, {\pid #1}.\mathit{#2} \, \m{then} \, #3 \, \m{else} \, #4}
\newcommand{\gencond}{\cond pe{C_1}{C_2}}
\newcommand{\call}[1]{#1}
\newcommand{\gencall}{\call X}
\newcommand{\rec}[2]{#1  =  #2}
\newcommand{\genrec}{\rec{X}{C_X}}
\newcommand{\chorterm}[2]{\m{def} \, #1 \, \m{in} \, #2}
\newcommand{\rname}[2]{\mbox{\small\textsc{\MakeLowercase{#1-#2}}}}
\newcommand{\eval}[4]{{#1}\downarrow^{#2}_{\pid{#3}}{#4}}
\newcommand{\lcond}[2]{\pid{#1}:\m{#2}}
\newcommand{\pn}{\m{pn}}
\newcommand{\precongr}[1][\procs]{\preceq_{#1}}
\newcommand{\epp}[2][]{[\![#2]\!]_{\pid{#1}}}
\renewcommand{\merge}{\sqcup}

\newcommand{\extract}[2][]{(\![{#2}]\!)_{#1}}
\newcommand{\rwto}{\leadsto}
\newcommand{\rwtot}{\leadsto^\ast}
\newcommand{\dlock}{\boldsymbol 1}
\newcommand{\mr}{\mathrel{\mathcal R}}
\newcommand{\bisim}{\sim}
\newcommand{\lthen}[2]{\pid{#1}.{#2}:\m{then}}
\newcommand{\lelse}[2]{\pid{#1}.{#2}:\m{else}}
\newcommand{\smallnet}[2][0cm]{\makebox[#1][c]{\ensuremath{#2}}}

\newcommand{\genmulticometa}{(\til \eta)}
\newcommand{\multicom}[1]{\left(\begin{array}c #1 \end{array}\right)}
\newcommand{\rcv}{\m{rcv}}
\newcommand{\snd}{\m{snd}}

\begin{document}
\pagestyle{plain}

\title{Implementing Choreography Extraction}
\author[1]{Lu\'\i s Cruz-Filipe}
\author[1]{Kim S. Larsen}
\author[1]{Fabrizio Montesi}
\author[2]{Larisa Safina}

\affil[1]{University of Southern Denmark \{lcf,kslarsen,fmontesi\}@imada.sdu.dk}
\affil[2]{INRIA, larisa.safina@inria.fr}
\date{}

\maketitle

\begin{abstract}
  Choreographies are global descriptions of interactions among concurrent components, most notably
  used in the settings of verification and synthesis of correct-by-construction software.
  They require a top-down approach: programmers first write choreographies, and then use them to
  verify or synthesize their programs.
  However, most software does not come with choreographies yet, which prevents their application.
  To attack this problem, previous work investigated choreography extraction, which automatically
  constructs a choreography that describes the behaviour of a given set of programs or protocol
  specifications.

  We propose a new extraction methodology that improves on the state of the art: we can deal with
  programs that are equipped with state and internal computation and time complexity is dramatically better.
  We also implement this theory and show that, in spite of its theoretical exponential complexity,
  it is usable in practice.
  We discuss the data structures needed for an efficient implementation, introduce some
  optimisations, and perform a systematic practical evaluation.
\end{abstract}

\section{Introduction}
\label{sec:intro}


The standard way of specifying the behaviour of a system of communicating processes is to describe their individual behaviours.
Some important questions about these systems are ``Is it free from deadlocks?'' and ``Is it free from livelocks?''. Answering these questions is undecidable in general.
To answer these these and other similar questions, we can study a more general problem: what does the system \emph{do}?
In particular, what are the communications that the system will enact?
In this paper, we develop an automatic procedure that answers this question and that is efficient enough in practice.

As an example, consider the following (pseudocode) specification of a simple single sign-on scenario inspired by the OpenID protocol~\cite{openid}.
It describes a network with three processes: a user (\pid u) tries to access a third-party web service (\pid w) by verifying their identity at an authentication service (\pid a). The processes interact by using primitives for sending and receiving values (\texttt{send} and \texttt{recv}), and choosing from and offering alternative behaviours (\texttt{choose} and \texttt{offer}).

\begin{center}
\ttfamily\small
\begin{tabular}{l|l|l}
  \toprule
  \multicolumn1c{\rmfamily Program for \pid u} & \multicolumn1c{\rmfamily Program for \pid a} & \multicolumn1c{\rmfamily Program for \pid w} \\
  \midrule
  procedure X:                 & procedure X:                            & procedure X: \\
  \quad send cred to \pid a   & \quad recv c from \pid u               & \quad offer to \pid a: \\
  \quad offer to \pid a:      & \quad if check(c):                      & \qquad OK: send t to \pid u \\
  \qquad OK: recv token from w & \qquad choose OK at \pid u             & \qquad KO: call X \\
  \qquad KO: call X            & \qquad choose OK at \pid w\\
							   & \quad else: \\
							   & \qquad choose KO at \pid u \\
                               & \qquad choose KO at \pid w\\
                               & \qquad call X \\
  call X                       & call X                                  & call X \\
  \bottomrule
\end{tabular}
\end{center}

To answer the question of what this system does, we can use \emph{choreographic languages}---languages that describe the behaviour of an entire system from a global viewpoint. Examples of such languages are Message Sequence Charts~\cite{msc}, the W3C Web Services Choreography Description Language~\cite{wscdl}, and the Business Process Modelling Notation~\cite{bpmn}.
In the language that we use in this article, the behaviour of the system above can be given as the following choreography.
\begin{align*}
\m{def}\ X ={} &
	\com{u}{cred}{a}{c};\\
	& \m{if}\ \pid a.check(c)\ \m{then}\ 
		\sel{a}{u}{ok};\
		\sel{a}{w}{ok};\
		\com{w}{t}{u}{token}\\
	& \phantom{\m{if}\ \pid a.check\ }\makebox[0em][l]{\m{else}}\phantom{\m{then}\ }
		\sel{a}{u}{ko};\
		\sel{a}{w}{ko};\
		X
	\\
\m{in}\ X\hspace*{1.8em}
\end{align*}
Here, \code{-\hspace{-0.3mm}>} denotes a communication from the left- to the right-hand process. This choreography describes the global protocol: the authentication service receives the user's credentials, and then decides whether the user should get a session token ($t$) from the web service, or reattempt authentication (by reinvoking procedure $X$).

The general problem of synthesising a representative choreography from a set of process specifications is called \emph{choreography extraction} (extraction for short)~\cite{CMS18}.
Extraction is a hard problem, since it requires predicting how concurrent processes can communicate with each other. Approaching this problem with brute force leads to the typical case explosion for static analysis of concurrent programs \cite{O18}.
Extraction is also connected to deadlock-freedom: any system that can be represented by a choreography is necessarily deadlock-free~\cite{CM20}; however, some systems are deadlock-free but cannot be extracted to a choreography~\cite{CLM17}.
The state-of-the-art implementation of extraction~\cite{LTY15} has worst-case super-factorial complexity. This limits the feasibility of thorough testing, and to date the practical limits of extraction are still largely unexplored and unclear.

\subsection*{Contribution}
In this article, we present a simple yet effective choreography extraction procedure, whose correctness and efficiency are systematically tested in practice.
We revisit and expand on the key ideas that we previously presented in~\cite{CLM17}, where we informally described an extraction algorithm. We formally define this algorithm and prove its main properties. In this process, we also made some small improvements and extensions.
Then, we introduce an implementation of our algorithm and carry out the first thorough and systematic testing of choreography extraction in the literature.

Our contribution is three-fold.

\paragraph{Theory.} Our theory for choreography extraction focuses on simplicity.
The languages for choreographies and processes respectively build upon Core Choreographies and Stateful Processes, which have been previously proposed as languages for foundational studies on choreographies: they are designed to be minimalistic, yet representative of the choreographic approach and Turing complete~\cite{CM20,CMP21a}.

We extend the process language with an abstract operational semantics that overapproximates the possible executions of a network (a system of processes).
This abstract semantics allows us to construct a finite graph that represents the (abstract) execution space of a system. Choreography extraction can then be formulated as a procedure that reconstructs a choreography by following paths in this graph.
Our extraction also helps in debugging: if a potential deadlock is present, we pinpoint it with a special term ($\dlock$). Choreographies that are successfully extracted guarantee deadlock-freedom. The soundness of our approach is proven in terms of strong bisimilarity~\cite{S11}.

Using our theory as foundation, we design an algorithm that is significantly simpler and more efficient than previous work: it consists of only two phases (the construction of the graph and its visit) and has better complexity.

\paragraph{Implementation}
The design of our implementation includes choosing adequate data structures, optimising substeps, parallelisation, and proving all these choices correct. An example is devising an efficient decision procedure for guaranteeing the absence of livelocks.
As a result, we obtain an implementation that successfully manages our test suite (described next) in reasonable time.

\paragraph{Evaluation}
Designing a test suite for extraction poses a major challenge: we cannot simply generate random networks since nearly none of them will be extractable (it is very unlikely that randomly-generated processes have matching communication actions throughout execution). In order to ensure that we generate extractable networks, we rely on a compilation procedure for choreographies that has been proven formally correct~\cite{CMP21b}.

Specifically, by generating choreographies and compiling them, we obtain a first set of networks that are guaranteed to be extractable. This set is then extended to a more comprehensive test suite by applying additional transformations that simulate realistic software development:
we devised an automatic tool that simulates the typical changes (both correct and incorrect) that are introduced when a programmer edits a local process program, and then tried to extract choreographies from the edited networks. This provides information on how quickly our program fails for unextractable networks.

Our test suite represents the first systematic and comprehensive approach to the evaluation of extraction. Thus, we believe it to be a useful reference also for the future design and implementations of new extraction algorithms.

\subsection{Related Work}
\label{sec:related}

Most works on choreographic languages focus on the inverse (and simpler) operation to extraction: Endpoint Projection (EPP), the translation of choreographies into distributed implementations~\cite{CHY12,Hetal16}.
EPP supports a top-down development methodology: developers first write choreographies and then execute the output mechanically generated by EPP. However, there are scenarios
where this methodology is not applicable:
\begin{itemize}
\item The analysis or use of ``legacy code'', i.e., code that was not generated by EPP.
This might be code that was developed previously, or new code written in a technology that does not adopt EPP.
With legacy code, EPP is not helpful.
\item Code updates: the programs generated by EPP are typically updated locally later on (for
configuration or optimisations, for example). Since the original choreography is not automatically 
updated, rerunning EPP loses these changes. Also, we lose the information on what the system is actually doing, since the original choreography does not represent it anymore.
\end{itemize}

To attack these issues, researchers started investigating choreography extraction, which is the topic of this article~\cite{CMS18,LT12,LTY15}.
Extraction still represents a green field of research. Early attempts developed theories based on session types~\cite{LT12}, linear logic~\cite{CMS18}, or communicating automata~\cite{LTY15}.
The theory in~\cite{LTY15} comes with an implementation, which is the state of the art in the area. However, the proposed algorithm does not focus on efficiency, nor simplicity: it consists of several complex phases, one of which has worst-case super-factorial complexity.
This limits the feasibility of thorough testing, and indeed the implementation has been tested on small selected examples, which tell us little about its applicability on a larger scale and its correctness.
The choreographic language in~\cite{LTY15} is different than ours, for example it cannot capture internal computation and it has internal threads (which we represent as separate processes). Nevertheless, many of the manually-written examples in~\cite{LTY15} can be reformulated in our framework. These reformulations are included in the testing of our implementation, and some of them benefit greatly from our parallelisation of extraction.

\section{Networks, Choreographies, and Extraction}
\label{sec:theory}
We introduce the languages that we use in this work to model process networks and choreographies.
These languages are very similar to those studied in~\cite{CM20}, where the interested reader can
find a formal treatment of these calculi, as well as statements and proofs of the most relevant
properties.

Syntactically, there are minor differences due to the goal of obtaining a process language closer to
real implementation languages, as depicted by the example in the introduction.
These changes are inspired by the languages discussed in~\cite{CM17a}.
Semantically, the reduction semantics for these calculi is also extended with labels in order to
allow for a formalisation of the link between choreographies and their process implementations as a
bisimilarity.

\subsection{Networks}
\label{sec:networks}

Process networks, or simply networks, represent systems of concurrent communicating processes.
Each process has an internal memory where values can be stored, identified by
variables.
Our model of networks is a calculus, which we call \emph{Stateful Processes} (SP), parameterised  on
sets of process names, expressions, labels, variables, and procedure names.
We assume these sets to be fixed, as they are immaterial for our presentation.
We abstract from the concrete language of expressions, which models internal computation and is
orthogonal to our development, assuming only that: expressions can contain values $v$ and variables;
and evaluation of expressions always terminates and returns a value.

To simplify the presentation, we use $\pid p,\pid q,\ldots$ to range over process
names, $e,e',e_1,\ldots$ to range over expressions, $\ell,\ell',\ell_1,\ldots$ to range over labels,
$x,x',y,y_1,\ldots$ to range over variables, and $X,Y,X',Y_1,\ldots$ to range over procedure names.
Networks are ranged over by $N,N',N_1,M,\ldots$.

\paragraph{Syntax.}
Formally, a network is a map from a finite set of process names to \emph{processes} of the form $\procterm{\{\brec{X_i}{B_i}\}_{i\in I}}{B}$, where each $B_i$ and $B$ are process \emph{behaviours}.
We denote by $\actor{p_1}{P_1}\parp\cdots\parp\actor{p_n}{P_n}$ the network $N$ that maps each
process name $\pid p_i$ to the process term $P_i$, i.e., $N(\pid p_i)=P_i$ for all $i\in[1,n]$, and
every other process name to $\nil$.
Note that the order of processes in this representation is immaterial.
The network mapping all process names to $\nil$ is denoted $\nil$.

In $\procterm{\{\brec{X_i}{B_i}\}_{i\in I}}{B}$, behaviour $B$ is the \emph{main behaviour} of the process, and $\{\brec{X_i}{B_i}\}_{i\in I}$ is a
set of \emph{procedure definitions}, assigning each $X_i$ to the corresponding behaviour $B_i$ (the
\emph{body} of the procedure).
Behaviours are syntactically defined by the grammar in Figure~\ref{fig:sp_syntax}.
\begin{figure}
  \centering
  \[B ::= \nil \mid \gencall \mid \bsend qe;B \mid \brecv px;B \mid \bsel q\ell;B
  \mid\bbranch p{\ell_1:B_1,\ldots,\ell_n:B_n} \mid \bcond e{B_1}{B_2}
  \]
  \caption{Syntax of Stateful Processes.}
  \label{fig:sp_syntax}
\end{figure}
We use $P,P',P_1,\ldots$ to range over processes and $B,B',B_1,\ldots$ to range over behaviours.
We also write $\procs[p]$ for the set of all procedure definitions at $\pid p$, and we often
abbreviate $\actor p{\procterm{\procs[p]}B}$ to $\actor[{\procs[p]}]pB$, or simply $\actor pB$ if $\procs[p]$ is clear from the context.

Term $\nil$ is the behaviour of a process that has terminated.

Term $X$ is a procedure call, i.e., the invocation of the procedure called $X$ in the process
executing the behaviour.
Procedure calls are executed by replacing them with their definition.

Term $\bsend qe;B$ is a send action, which evaluates expression $e$, sends the resulting value to
process $\pid q$, and continues as $B$.
Dually, term $\brecv px;B$ receives a value from process $\pid p$, stores it in a local variable
$x$, and continues as $B$.

Term $\bsel q\ell;B$ sends to $\pid q$ the selection of a behaviour labelled by $\ell$ (labels are
constants), and then proceeds as $B$.
Selections are received by the branching term $\bbranch p{\ell_1:B_1,\ldots,\ell_n:B_n}$, which
models the offering of different possible behaviours: the process executing this term waits to
receive from $\pid p$ the selection of one of the labels $\ell_i$ in $\ell_1,\ldots,\ell_n$, and
then proceeds with the associated behaviour $B_i$.

Term $\bcond e{B_1}{B_2}$ is the standard conditional term.
It evaluates the Boolean expression $e$ and proceeds as $B_1$ if the result is \m{true}, and as
$B_2$ otherwise.

Networks are expected to satisfy some well-formedness conditions, corresponding to usual
requirements in practice:
\begin{itemize}
\item processes do not contain subterms that attempt self-communication (for example, $\actor p{\brecv px}$
  is not allowed);
\item all expressions in guards of conditionals evaluate to \m{true} or \m{false};
\item all procedure calls refer to procedures defined in the enclosing process;
\item all defined procedures are distinct, i.e., in $\procterm{\{\rec{X_i}{B_i}\}_{i \in I}}{B}$,
  $X_i\neq X_j$ for every $i\neq j\in I$.
\end{itemize}
Note that we do \emph{not} require procedure calls to be guarded.

\begin{example}\label{ex:auth_sp}
The example network from the introduction can be formalised as follows.
\begin{align*}
  & \actor u{
    \procterm
        {\brec X{\bsend a{cred};\bbranch a{\mathit{ok}: \brecv w{token},\ \mathit{ko}: \call X}}
          \\ &\hspace*{1.5em}}
        {\call X}}
  \\
  \parp \
  & \actor a{
    \procterm
        {\brec X{\brecv u{cred};\bcond{check(cred)}
            {\left( \bsel u{ok}; \bsel w{ok} \right)}
            {\left( \bsel u{ko}; \bsel w{ko}; \call X \right)}}
          \\ &\hspace*{1.5em}}
        {\call X}}
  \\
  \parp \
  & \actor w{
    \procterm
        {\brec X{\bbranch a{\mathit{ok}: \bsend u{token},\ \mathit{ko}: \call X }}
          \\ &\hspace*{1.5em}}
        {\call X}}
\end{align*}
This corresponds precisely to the example written earlier, but now using the formal language of SP.
We follow the usual practice of omitting trailing $\nil$ terms in behaviours.
\eoe
\end{example}

The inductive definition of process behaviours gives rise to a notion of context in the usual
way~\cite{SW01}, by allowing the terminal $\nil$ to be replaced by a hole.

\paragraph{Semantics.}
The semantics of SP is given in terms of labelled reductions of the form
$N,\sigma\lto\lambda N',\sigma'$, where $\sigma$ is a \emph{state function} (which, given a
process and a variable, returns the value stored in that variable in the process's memory) and
$\lambda$ is a \emph{reduction label}.
The syntax of reduction labels is given in Figure~\ref{fig:cc_reductions}.
\begin{figure}
  \[
  \lambda ::= \lcom pvq \mid \gensel \mid \lcond p{then} \mid \lcond p{else}
  \]
  \caption{Reduction labels for reductions in Stateful Processes}
  \label{fig:cc_reductions}
\end{figure}
The role of labels is to identify the action that has been performed; this will be useful later to
state and prove results about the extraction algorithm.

In realistic implementations, the state of each process's memory would be stored locally with each process. The formulation chosen here is trivially equivalent, but using a global state function simplifies
the formulation of some of our later results, as in other works on choreographies~\cite{CMP21a,CMP21b}.

The semantics of SP is defined by the rules in Figure~\ref{fig:sp_semantics}.
\begin{figure}
  \begin{eqnarray*}
    &\infer[\rname{S}{Com}]{
      \actor[{\procs[p]}]p{\bsend qe;B_1} \parp
      \actor[{\procs[q]}]q{\brecv px;B_2},\sigma
      \lto[]{\lcom pvq}
      \actor[{\procs[p]}]p{B_1} \parp
      \actor[{\procs[q]}]q{B_2},\sigma[\tuple{\pid q,x} \mapsto v]
    }{
      \eval e\sigma pv
    }
    \\[1ex]
    &\infer[\rname{S}{Sel}]{
      \actor[{\procs[p]}]p{\bsel q{\ell_j};B} \parp
      \actor[{\procs[q]}]q{\bbranch p{\ell_1:B_1,\ldots,\ell_n:B_n}},\sigma
      \lto[]\gensel
      \actor[{\procs[p]}]pB \parp \actor[{\procs[q]}]q{B_j},\sigma
    }{
      1\leq j\leq n
    }
    \\[1ex]
    &\infer[\rname{S}{Then}]{
      \actor[{\procs[p]}]p{\bcond e{B_1}{B_2}},\sigma
      \lto[]{\lcond p{then}}
      \actor[{\procs[p]}]p{B_1},\sigma
    }{
      \eval e\sigma p{\m{true}}
    }
    \\[1ex]
    &\infer[\rname{S}{Else}]{
      \actor[{\procs[p]}]p{\bcond e{B_1}{B_2}},\sigma
      \lto[]{\lcond p{else}}
      \actor[{\procs[p]}]p{B_2},\sigma
    }{
      \eval e\sigma p{\m{false}}
    }
    \\[1ex]
    &
    \infer[\rname{S}{Par}]{
      N \parp M,\sigma \lto[]\lambda N' \parp M,\sigma'
    }{
      N,\sigma \lto[]\lambda N',\sigma'
    }
    \qquad
    \infer[\rname{S}{Struct}]{
      N,\sigma \lto[]\lambda N', \sigma'
    }{
      N \precongr[] M
      & M,\sigma \lto[]\lambda M',\sigma'
      & M' \precongr[] N'
    }
  \end{eqnarray*}
  \caption{Semantics of Stateful Processes.}
  \label{fig:sp_semantics}
\end{figure}
Two processes can synchronise when they refer to each other.
In rule~\rname{S}{Com}, an output at $\pid p$ directed at $\pid q$ synchronises with the dual input
action at $\pid q$ -- intention to receive from $\pid p$.
The communicated value ($v$) is obtained by evaluating expression $e$ locally at the sender $\pid p$ taking into account the memory state $\sigma$, denoted $\eval e\sigma pv$, and stored in the corresponding variable $x$ at $\pid q$ in the reductum.
The label in the reduction summarises the observable part of the communication.

Rule~\rname{S}{Sel} follows the same intuition, but for a label selection -- where $\pid p$ selects
between different possible behaviours offered at $\pid q$ by sending the appropriate label.

Rules~\rname{S}{Then} and~\rname{S}{Else} model conditionals in the expected way, while rule
\rname{S}{Par} allows for reductions involving only a subset of processes in the network.

Rule~\rname{S}{Struct} closes reductions under a \emph{structural precongruence relation}
$\precongr[]$, generated by closing the rule in Figure~\ref{fig:sp_precongr} under reflexivity,
transitivity, and context.
\begin{figure}
  \begin{eqnarray*}
    \infer[\rname{S}{Unfold}]{
      \actor[{\procs[p]}]p{\call X} \precongr[] \actor[{\procs[p]}]p{B_X}
    }{
      \rec{X}{B_X}\in\procs[p]
    }
  \end{eqnarray*}
  \caption{Structural precongruence in Stateful Processes.}
  \label{fig:sp_precongr}
\end{figure}
This rule allows procedure calls to be replaced by their definition anywhere inside a process's
behaviour.

\begin{lemma}[Determinism of SP]
  \label{lem:sp-confluent}
  Let $N$ be a network, $\sigma$ be a state, and $\lambda$ be a reduction label.
  For any networks $N_1$ and $N_2$ and states $\sigma_1$ and $\sigma_2$, if
  $N,\sigma\lto\lambda N_i,\sigma_i$ for each $i=1,2$, then $\sigma_1=\sigma_2$ and there exists a
  network $N'$ such that $N_i\precongr[]N'$ for $i=1,2$.
\end{lemma}
\begin{proof}[Proof (sketch).]
  First observe that labels uniquely identify the process(es) involved in the reduction: this is
  trivially the case for rules \rname{S}{Com}, \rname{S}{Sel}, \rname{S}{Then} and \rname{S}{Else},
  and rules \rname{S}{Par} and \rname{S}{Struct} preserve this property.

  Furthermore, the action(s) being executed must be the head action(s) in each participating
  process, possibly after unfolding a behaviour consisting of a procedure call: once again, this is
  trivially the case for the rules that execute reductions, and preserved by \rname{S}{Par} and
  \rname{S}{Struct} (in the latter case, because structural congruence cannot change the head action
  of a process unless it is a procedure call).
  
  Therefore the label of the reduction uniquely determines the resulting state; and the resulting
  networks may differ only in the procedure calls that have been unfolded in each process.
  If $N,\sigma\lto[]\lambda N_1,\sigma_1$ and $N,\sigma\lto[]\lambda N_2,\sigma_2$, it thus follows
  that $\sigma_1=\sigma_2$, and that $N_1\precongr[]N'$ and $N_2\precongr[]N'$ for the network $N'$
  obtained from $N_1$ by unfolding all procedure calls that have been unfolded in $N_2$ and vice versa.
\end{proof}

\subsection{Core Choreographies}

Networks define the local actions that each process should perform, as in Example~\ref{ex:auth_sp},
but they can be hard to read and error-prone to write: each process can have a different structure,
because they carry out interactions with different other processes at different times, yet we must
ensure that each action aiming at interacting with another process is going to be matched eventually
by a compatible action at that process.
Conversely, choreographies are specifications on a higher level of abstraction that make the flow of
interactions easy to read and write, instead of focusing on the local view of each process.

We express choreographies using a minimalistic formal language (but still expressive enough to
capture relevant practical examples from the literature, as we show later).
Like networks, choreographies range over sets of process names, expressions, labels, and procedure
names, with the same conventions as above.
We call the choreography language in this work Core Choreographies (CC).

\paragraph{Syntax.}
A choreography is a term of the form $\chorterm{\{\rec{X_i}{C_i}\}_{i \in I}}{C}$, where: $C$ is the
\emph{main body} of the choreography; and $\{\rec{X_i}{C_i}\}_{i\in I}$ is a set of procedure
definitions, mapping each recursion variable $X_i$ (the \emph{name} of the procedure) to the
respective choreography \emph{body} $C_i$, for some finite set of indices $I$.
Choreography bodies are defined inductively by the grammar in Figure~\ref{fig:cc_syntax}.
\begin{figure}
  \[ C ::= \nil \mid \gencom; C \mid \gensel; C \mid \gencond \mid \gencall \]
  \caption{Syntax of Core Choreographies}
  \label{fig:cc_syntax}
\end{figure}
We often abuse terminology and refer to choreography bodies as ``choreographies'', when no confusion
can arise.
 
Term $\nil$ is the terminated choreography, and again we typically omit it in examples when it is
the trailing term of a non-terminated choreography.

The next two terms both model systems that execute an interaction and proceed as $C$.
There are two kinds of interactions.
\begin{itemize}
\item In a value communication $\gencom$, process $\pid p$ evaluates expression $e$ and sends the
  result to process $\pid q$, which stores it in its variable $x$,
  replacing the value previously stored there.
  We abstract from the concrete language of expressions $e$, which models internal computation and
  is orthogonal to our development, assuming only that: expressions can contain values $v$ and
  variables; and evaluation of expressions always terminates and returns a value.
\item In a selection $\gensel$, $\pid p$ selects $\ell$ among the set of branches offered by
  $\pid q$.
\end{itemize}
We use $\eta$ to range over interactions, when we do not need to distinguish between value
communications and label selections.


In a conditional $\gencond$, $\pid p$ evaluates the (Boolean) expression $e$ and checks whether the
result is \m{true} or \m{false} to decide whether the system proceeds as $C_1$ or $C_2$,
respectively.

Finally, term $\gencall$ is a procedure call.
Intuitively, executing $\gencall$ corresponds to executing the body of the procedure with name $X$.

As for networks, we often omit the first part of choreography terms that have the empty set as the
set of procedure definitions, i.e., we simply write $C$ instead of $\chorterm{\emptyset}{C}$.

We assume that all choreographies are well-formed, meaning that:
\begin{itemize}
\item there are no self-communications, i.e., $\pid p$ and $\pid q$ are distinct in every subterm of
  the form $\gencom$ or $\gensel$;
\item all expressions in guards of conditionals evaluate to \m{true} or \m{false};
\item all procedure calls are guarded in procedure definitions, i.e., there is no procedure
  definition of the form $X=Y$ for some variables $X$ and $Y$;
\item all defined procedures are distinct, i.e., in $\chorterm{\{\rec{X_i}{C_i}\}_{i \in I}}{C}$,
  $X_i\neq X_j$ for every $i\neq j\in I$.
\end{itemize}

As before, from the inductive definition of choreographies we define contexts in the usual
way~\cite{SW01}.

\paragraph{Semantics.}

The semantics of CC is given by labelled reductions $C,\sigma\lto\lambda C',\sigma'$, with labels
$\lambda$ as in SP.
The reduction rules are given in Figure~\ref{fig:cc_semantics}.
\begin{figure}
  \begin{eqnarray*}
    &\infer[\rname{C}{Com}]{
      \gencom;C,\sigma \lto{\lcom pvq} C,\sigma[\tuple{\pid q,x} \mapsto v]
    }{
      \eval e\sigma pv
    }
    \qquad
    \infer[\rname{C}{Sel}]{
      \gensel;C,\sigma \lto\gensel C,\sigma
    }{}
    \\[1ex]
    &\infer[\rname{C}{Then}]{
      \gencond,\sigma \lto{\lcond p{then}} C_1, \sigma
    }{
      \eval e\sigma p{\m{true}}
    }
    \quad
    \infer[\rname{C}{Else}]{
      \gencond,\sigma \lto{\lcond p{else}} C_2, \sigma
    }{
      \eval e\sigma p{\m{false}}
    }
    \\[1ex]
    &
    \infer[\rname{C}{Struct}]{
      C_1,\sigma \lto\lambda C'_1,\sigma'
    }{
      C_1 \precongr C_2
      & C_2,\sigma \lto\lambda C'_2, \sigma'
      & C'_2 \precongr C'_1
    }
  \end{eqnarray*}
  \caption{Semantics of Core Choreographies}
\label{fig:cc_semantics}
\end{figure}
The first four rules formalise the above informal description of the involved syntactic terms, and follow the same intuitions as the corresponding rules for SP.

Rule~\rname{C}{Struct} closes reductions under a structural precongruence $\precongr$ that allows
procedure calls to be unfolded and non-interfering actions to be executed in any order.
The main rules defining this relation are given in Figure~\ref{fig:cc_precongr}; the missing rules
close this relation under reflexivity, transitivity, and context.
\begin{figure}
  \begin{eqnarray*}
    &\infer[\rname{C}{Eta-Eta}]{
      \eta;\eta';C \precongr \eta';\eta;C
    }{
      \pn(\eta)\cap\pn(\eta') = \emptyset
    }
    \\[1ex]
    &\infer[\rname{C}{Eta-Cond}]{
      \cond pe{(\eta;C_1)}{(\eta;C_2)}
      \precongr 
      \eta;\gencond
    }{
      \pid p\notin \pn(\eta)
    }
    \\[1ex]
    &\infer[\rname{C}{Cond-Eta}]{
      \eta;\gencond
      \precongr 
      \cond pe{(\eta;C_1)}{(\eta;C_2)}
    }{
      \pid p\notin \pn(\eta)
    }
    \\[1ex]
    &\infer[\rname{C}{Cond-Cond}]{
      \begin{array}{c}
	\cond pe{(\cond q{e'}{C_1}{C_2})}{(\cond q{e'}{C'_1}{C'_2})}
	\\
	\precongr
	\\
	\cond q{e'}{(\cond pe{C_1}{C'_1})}{(\cond pe{C_2}{C'_2})}
      \end{array}
    }{
      \pid p \neq \pid q
    }
    \\[1ex]
    &\infer[\rname{C}{Unfold}]{
      \gencall \precongr C_X
    }{
      \genrec\in\procs}
  \end{eqnarray*}
  \caption{Structural precongruence in Core Choreographies}
  \label{fig:cc_precongr}
\end{figure}

The key idea behind $\precongr$ is illustrated by rule~\rname{C}{Eta-Eta}, which swaps
communications between disjoint sets of processes (modeling concurrency).
In this rule, $\pn(C)$ denotes the set of process names that appear in $C$.
Rules~\rname{C}{Eta-Cond} and \rname{C}{Cond-Cond} are similar, as well as rule~\rname{C}{Cond-Eta}, which is dual to~\rname{C}{Eta-Cond}.
Rule~\rname{C}{Unfold} allows procedure calls to be replaced by the corresponding definition.

Since all rules except for~\rname{C}{Unfold} are reversible, one is often working with
choreographies $C_1$ and $C_2$ such that $C_1\precongr C_2$ and $C_2\precongr C_1$.
In this case, we write simply $C_1\equiv C_2$.

\begin{example}\label{ex:auth_cc}
  We can write the client authentication protocol in the introduction as a choreography in the
  following way, where $\sel a{u,w}\ell$ is a shortcut for $\sel au\ell;\sel aw\ell$.
  For presentation purposes, we write each procedure definition as a separate equation, and abuse notation by identifying the main body with procedure $\m{main}$.
  \begin{align*}
    X = {} & \com u{pwd}ax;\\
    & \cond a{\m{ok}(x)}{\left(\sel a{u,w}{ok};\com wtux\right)\\
      &\hspace{4em}}{\left(\sel a{u,w}{ko};X\right)} \\
    \m{main} = {} & X
  \end{align*}
  Here, $\pid u$ sends a password to $\pid a$.
  If this password is correct, $\pid a$ notifies $\pid u$ and $\pid w$, and $\pid w$ sends an
  authentication token $t$ to $\pid u$.
  Otherwise, $\pid a$ notifies $\pid u$ and $\pid w$ that authentication failed, and a new attempt
  is made (by recursively invoking $X$).

  This choreography can be obtained from the network in Example~\ref{ex:auth_sp} by the extraction
  algorithm defined in later sections.
  \eoe
\end{example}

\subsection{EndPoint Projection}

Choreographies satisfying some realisability conditions can be translated automatically into
networks by a transformation known as \emph{EndPoint Projection} (EPP).
We summarise this procedure, as it is a key ingredient to stating soundness of extraction.

Intuitively, EPP is defined by translating each choreography action into its local counterparts.
For example, a communication action $\gencom$ is projected as $\bsend qe$ for process $\pid p$, as
$\brecv px$ for process $\pid q$, and as a no-op for any other process.
The interesting case (where realisability plays a role) is the case of conditionals: for any process
other than $\pid p$, $\gencond$ must be projected as a unique behaviour.
This is dealt with by a partial operator called \emph{merging}~\cite{CHY12,CMP21b}.
Two behaviours are mergeable if every place where they differ is protected by a label selection.

The key rule defining merge is that for branching terms:
\begin{multline*}
  \bbranch p{\ell_i:B_i\mid i\in J} \merge
  \bbranch p{\ell_i:B'_i \mid i\in K} =\\
  \pid p\&\left(\{\ell_i:(B_i \merge B'_i)\mid i\in J\cap K\}
  \cup\{\ell_i:B_i\mid i\in J \setminus K\}\cup\{\ell_i:B'_i \mid i\in K \setminus J\}\right)
\end{multline*}
The remaining rules extend this operator homomorphically,
e.g., $(\bsend qe;B_1)\merge(\bsend qe;B_2)=\bsend qe;(B_1\merge B_2)$.
Merge is undefined for two behaviours that start with different actions, e.g., $\bsend qe;B$ and
$\brecv qx;B'$.

The EPP of a choreography body for process $\pid r$, denoted $\epp[r]C$, is defined in
Figure~\ref{fig:epp}.
This extends to a choreography $\chorterm{\{\rec{X_i}{C_i}\}_{i \in I}}{C}$ by defining, for each
$\pid r$, $\procs[r]=\{\brec{X_i}{\epp[r]{C_i}}\}_{i\in I}$ and $\epp C$ as the function mapping
each process $\pid r$ to $\procterm{\procs[r]}{\epp[r]C}$.
If this is defined for every $\pid r$, the choreography is said to be
\emph{projectable}.\footnote{In practice, some static analysis is performed to optimise the projection of procedure invocations so that $\epp[r]{X} = \nil$ if $\pid r$ cannot be involved in the execution of $X$~\cite{CM20}. This optimisation does not affect our results.}

\begin{figure}
\begin{align*}
&\epp[r]{\gencom;C} =
\begin{cases}
  \bsend qe;\epp[r]C & \mbox{if $\pid r = \pid p$} \\
  \brecv px;\epp[r]C & \mbox{if $\pid r = \pid q$} \\
  \epp[r]C & \mbox{o.w.}
\end{cases}
\quad\qquad
\epp[r]{\gensel;C} =
\begin{cases}
  \bsel q\ell;\epp[r]C & \mbox{if $\pid r = \pid p$} \\
  \bbranch p{\ell:\epp[r]C} & \mbox{if $\pid r = \pid q$} \\
  \epp[r]{C} & \mbox{o.w.}
\end{cases}
\\[1ex]
&\epp[r]{\gencond} =
\begin{cases}
  \bcond e{\epp[r]{C_1}}{\epp[r]{C_2}} & \mbox{if $\pid r = \pid p$} \\
  \epp[r]{C_1} \merge \epp[r]{C_2} & \mbox{o.w.}
\end{cases}
\qquad
\epp[r]\nil = \nil
\qquad\epp[r]{X} = X
\end{align*}
\caption{Endpoint Projection of choreography bodies.}
\label{fig:epp}
\end{figure}

\begin{example}
  The network in Example~\ref{ex:auth_sp} is the EPP of the choreography in Example~\ref{ex:auth_cc}.
\end{example}

\section{Extraction from SP}
\label{sec:extraction}

In this section, we develop the theory of extracting a choreography from a network.
In a nutshell, the idea is to execute the network symbolically (abstracting from the actual
values that are communicated, for example) and use the trace of the execution to write down a choreography.
Since network reduction is non-deterministic and networks may have infinite behaviour, this poses
some challenges even to ensure termination.

We divide this presentation in two parts.
First, we focus on the fragment of SP without recursive definitions, which we use to discuss the
intuition behind our extraction algorithm in a simple setting.
We present extraction for this fragment, and formally state and prove its soundness.
In the second part, we extend the construction to deal with infinite behaviour.

\subsection{The finite case}
\label{sec:finite}
 
In this section we focus on \emph{finite SP}, the fragment of SP without recursive definitions.
Formally, networks in SP are well-formed networks where all processes are of the form
$\procterm\emptyset B$; in particular, $B$ cannot contain any procedure calls.

We start by formalising our intuitive notion of ``executing a network symbolically'' by means of a
rewriting relation over a language of extended choreography bodies.
\begin{definition}
  An \emph{extended choreography body} is a term written in the grammar of
  Figure~\ref{fig:cc_syntax} using the additional constructs $\extract{N}$, where $N$ is a network
  in finite SP, and $\dlock$, which stands for a deadlocked system.
\end{definition}

\begin{definition}
  \label{defn:extraction}
  We generate a rewriting relation $\rwto$ on extended choreography bodies by the
  rules
  \begin{align*}
    \extract \nil
    & \rwto \nil
    \\
    \extract{\actor[]p{\bsend qe;B_{\pid p}} \parp \actor[]q{\brecv px;B_{\pid q}} \parp N'}
    & \rwto\gencom;\extract{\actor[]p B_{\pid p} \parp \actor[]q B_{\pid q} \parp N'}
    \\
    \extract{
      \actor[]p{\bsel q{\ell_k};B_{\pid p}} \parp
      \actor[]q{\bbranch p{\ell_1:B_{\pid q_1},\ldots,\ell_n:B_{\pid q_n}}} \parp N'}
    & \rwto\sel pq{\ell_k};\extract{\actor[]p B_{\pid p} \parp \actor[]q B_{\pid q_k} \parp N'}
    \\
    \extract{\actor[]p{\bcond e{B_1}{B_2}}\parp N'}
    & \rwto\cond pe{\extract{\actor[]p B_1\parp N'}}{\extract{\actor[]p B_2\parp N'}} \\
    \extract N & \rwto\dlock\mbox{, if no other rule applies}
  \end{align*}
  closed under choreography contexts.
\end{definition}

Extraction operates by finding an action or a pair of matching actions in a network and replacing them
by the corresponding choreography action.
In general, there may be different options for these choices, making extraction nondeterministic.

\begin{example}
  \label{ex:non-deterministic}
  We illustrate this rewriting system with three example networks.
  \begin{itemize}
  \item Consider the network $N_1$ defined as 
    $\actor[] p{\bsend qe} \parp\actor[]q{\brecv px}
    \parp\actor[]r{\bsend s{e'}} \parp\actor[]s{\brecv ry}$.
    There are two sequences of extraction steps from $N_1$, namely
    \begin{align*}
      \extract{N_1}
      & \rwto \gencom;\extract{\actor[]r{\bsend s{e'}} \parp\actor[]s{\brecv ry}} \\
      & \rwto \gencom;\com r{e'}sy;\extract\nil \\
      & \rwto \gencom;\com r{e'}sy;\nil \\
      \mbox{and }\extract{N_1}
      & \rwto \com r{e'}sy;\extract{\actor[]p{\bsend qe} \parp\actor[]q{\brecv px}} \\
      & \rwto \com r{e'}sy;\gencom;\extract\nil \\
      & \rwto \com r{e'}sy;\gencom;\nil
    \end{align*}
    Observe that the resulting choreographies can be rewritten into each other by
    Rule~\rname{C}{Eta-Eta} (Figure~\ref{fig:cc_semantics}).
  \item Consider now the network $N_2$ defined as
    $\actor[]p{B_p} \parp \actor[]q{B_q}$, where
    $B_p=\bcond e{\bsel q\lleft;\bsend q1}{\bsel q\lright;\brecv qx}$ and
    $B_q=\bbranch p{\lleft:\brecv py,\ \lright:\bsend p2}$.
    The only sequence of extraction steps from $N_2$ is
    \begin{align*}
      \extract{N_2}
      & \rwto \cond pe
              {\extract{\actor[]p{\bsel q\lleft;\bsend q1} \parp \actor[]q{B_q}}}
              {\extract{\actor[]p{\bsel q\lright;\brecv qx} \parp \actor[]q{B_q}}} \\
      & \rwto \cond pe
              {\sel pq\lleft;\extract{\actor[]p{\bsend q1} \parp \actor[]q{\brecv py}}}
              {\sel pq\lright;\extract{\actor[]p{\brecv qx} \parp \actor[]q{\bsend p2}}} \\
      & \rwto \cond pe
              {\left(\sel pq\lleft;\com p1qy;\extract\nil\right)}
              {\left(\sel pq\lright;\com q2px;\extract\nil\right)} \\
      & \rwto \cond pe
              {\left(\sel pq\lleft;\com p1qy;\nil\right)}
              {\left(\sel pq\lright;\com q2px;\nil\right)}
    \end{align*}
  \item We now introduce an example involving the deadlocked term.
    Consider the network $N_3$ defined as
    $\actor[]p{\bsend q1;\bsend r2}
    \parp\actor[]q{\brecv px;\bsend r3}
    \parp\actor[]r{\bcond e{\brecv py}{\brecv qy}}$.
    Again, there are two possible sequences of extraction steps from $N_3$, but both include a
    deadlocked term in the result.
    \begin{align*}
      \extract{N_3}
      & \rwto \com p1qx;\extract{
        \actor[]p{\bsend r2}\parp\actor[]q{\bsend r3}\parp\actor[]r{\bcond e{\brecv py}{\brecv qy}}
      } \\
      & \rwto \com p1qx;\cond re
              {\extract{\actor[]p{\bsend r2}\parp\actor[]q{\bsend r3}\parp\actor[]r{\brecv py}}}
              {\extract{\actor[]p{\bsend r2}\parp\actor[]q{\bsend r3}\parp\actor[]r{\brecv qy}}} \\
      & \rwto \com p1qx;\cond re
              {\left(\com p2ry;\extract{\actor[]q{\bsend r3}}\right)}
              {\left(\com q3ry;\extract{\actor[]p{\bsend r2}}\right)} \\
      & \rwto \com p1qx;\cond re{\left(\com p2ry;\dlock\right)}{\left(\com q3ry;\dlock\right)}
    \end{align*}
    Alternatively, we can first rewrite the conditional on $\pid r$.
    \begin{align*}
      \extract{N_3}
      & \rwto \cond re
              {\extract{\actor[]p{\bsend q1;\bsend r2}\parp\actor[]q{\brecv px;\bsend r3}\parp\actor[]r{\brecv py}}}
              {\extract{\actor[]p{\bsend q1;\bsend r2}\parp\actor[]q{\brecv px;\bsend r3}\parp\actor[]r{\brecv qy}}} \\
      & \rwto \cond re
              {\left(\com p1qx;\extract{\actor[]p{\bsend r2}\parp\actor[]q{\bsend r3}\parp\actor[]r{\brecv py}}\right)}
              {\left(\com p1qx;\extract{\actor[]p{\bsend r2}\parp\actor[]q{\bsend r3}\parp\actor[]r{\brecv qy}}\right)} \\
      & \rwto \cond re
              {\left(\com p1qx;\com p2ry;\extract{\actor[]q{\bsend r3}}\right)}
              {\left(\com p1qx;\com q3ry;\extract{\actor[]p{\bsend r2}}\right)} \\
      & \rwto \cond re{\left(\com p1qx;\com p2ry;\dlock\right)}{\left(\com p1qx;\com q3ry;\dlock\right)}
    \end{align*}
    Note that the resulting extended choreographies can again be rewritten into each other (using
    rules~\rname{C}{Cond-Eta} and \rname{C}{Eta-Cond}).\eoe
  \end{itemize}
\end{example}

Indeed, non-determinism of extraction is of no practical consequence.

\begin{lemma}
  \label{lem:sound-fin}
  If $\extract N \rwtot C_1$ and $\extract N \rwtot C_2$, then $C_1\equiv C_2$.
\end{lemma}
\begin{proof}
  This follows by induction from the fact that $\rwto$ has the diamond property.
  To see that this is the case, observe that, if $N\rwto T_1$ and $N\rwto T_2$ with $T_1\neq T_2$,
  then these two rewrites cannot use the first or last rules in the definition of $\rwto$, and the
  choreography actions introduced in $T_1$ and $T_2$ cannot share process names.
  Therefore we can continue the reduction from $T_1$ by adding the choreography action in $T_2$,
  obtaining $T'_1$, and we can continue the reduction from $T_2$ by adding the choreography action
  from $T_1$, obtaining $T'_2$.
  Then $T'_1$ and $T'_2$ differ only in the choreography actions at the top, which can be exchanged
  by one of the precongruence rules for CC, as in the previous example.
\end{proof}

The converse also holds: if $N$ can be extracted to a choreography, then it can be extracted to any
structurally congruent choreography.
\begin{lemma}
  \label{lem:extract-equiv}
  If $\extract N \rwtot C_1$ and $C_1\precongr[] C_2$, then $\extract N \rwtot C_2$.
\end{lemma}
\begin{proof}
  By induction on the derivation of $C_1\precongr[] C_2$.
  If this derivation consists of a single step, then it is an application of one of the rules in
  Figure~\ref{fig:cc_precongr}, and that rule cannot be \rname{C}{Unfold}.
  It follows immediately that the thesis holds.
  Otherwise the thesis follows immediately from the induction hypothesis.
\end{proof}

There is one important design option to consider when extracting a choreography from a process
implementation: what to do with actions that cannot be matched, i.e., processes that get stuck.
There are two alternatives: restrict extraction to lock-free networks (networks where all processes
eventually progress, in the sense of~\cite{CDM14}), so that it becomes a partial relation;
or extract stuck processes to a new choreography term $\dlock$, with the same semantics as $\nil$.
We choose the latter option for debugging reasons.
Specifically, practical applications of extraction may annotate $\dlock$ with the code of the
deadlocked processes, giving the programmer a chance to see exactly where the system is unsafe, and
attempt at fixing it manually.
Better yet: since the code to unlock deadlocked processes in process calculi can be efficiently
synthesised~\cite{CDM14}, our method may be integrated with the technique in~\cite{CDM14} to suggest
an automatic system repair.

\begin{remark}
  If $\extract N \rwto C$ and $C$ does not contain $\dlock$, then $N$ is lock-free.
  However, even if $C$ contains $\dlock$, $N$ may still be lock-free: the code causing the deadlock
  may be dead code in a conditional branch that is never chosen during execution.
  Other kinds of liveness issues, e.g., livelocks and starvation, are not possible in finite SP, but will be relevant later when dealing with recursion.
\end{remark}

In order to relate a network with its extracted choreography, we use the standard notion of bisimilarity, noting that transition labels for choreographies and networks are the same.
\begin{definition}
A binary relation $\mr$ between choreographies and networks is a \emph{bisimulation} if:
\begin{itemize}
\item If $C \mr N$ and $C \lto[]\mu C'$, then there exists $N'$ such that $N \lto[]\mu N'$ and $C' \mr N'$.
\item If $C \mr N$ and $N \lto[]\mu N'$, then there exists $C'$ such that $C \lto[]\mu C'$ and $C' \mr N'$.
\end{itemize}

$C$ is \emph{bisimilar} to $N$, written $C \bisim N$, if there exists a bisimulation $\mr$ such that $C \mr N$.
\end{definition}

Extraction is sound: it yields a choreography that is bisimilar to the original network.
Also, for finite SP, it behaves as an inverse of EPP.
\begin{theorem}
  \label{thm:correct}
  Let $N$ be a finite SP. Then:
  \begin{enumerate}[(i)]
  \item If there exists a choreography $C$ such that $\extract N\rwto C$, then $C\bisim N$.
  \item If $N=\epp C$ for some choreography $C$, then $\extract N\rwto C$.
  \end{enumerate}
\end{theorem}
\begin{proof}\mbox{}
  \begin{enumerate}[(i)]
  \item We show that the relation $\mr$ defined by $C \mr N$ if $\extract N \rwto C$ is a bisimulation by induction on the size of $C$.
    We detail one representative case.

    Suppose that $C$ is $\gencom;C'$, whence $N$ is of the form
    $\actor[]p{\bsend qe;B_p}\parp\actor[]q{\brecv px;B_q}\parp N'$.

    The case when either $C$ or $N$ reduces by making a reduction labelled by $\lcom pvq$ is
    trivial, since both $C,\sigma\lto[\emptyset]{\lcom pvq}C',\sigma'$ and
    $N,\sigma\lto[]{\lcom pvq}\actor[]p{B_p}\parp\actor[]q{B_q}\parp N',\sigma'$ (assuming that
    $\eval e\sigma pv$), and the latter network extracts to $C'$

    Suppose that $C,\sigma\lto[\emptyset]\lambda C'',\sigma'$ for some other label $\lambda$.
    Due to the way structural congruence is defined, and since there are no procedure definitions,
    it follows that also $C',\sigma\lto[\emptyset]\lambda C''',\sigma'$, and that
    $C''\precongr[\emptyset]\gencom;C'''$.
    Since $\actor[]p{B_p}\parp\actor[]q{B_q}\parp N'\rwto C'$, by the induction hypothesis,
    $\actor[]p{B_p}\parp\actor[]q{B_q}\parp N',\sigma\lto[]\lambda\actor[]p{B_p}\parp\actor[]q{B_q}\parp N'',\sigma'$;
    but $\lambda$ cannot involve $\pid p$ or $\pid q$, so also
    $\actor[]p{\bsend qe;B_p}\parp\actor[]q{\brecv px;B_q}\parp N',\sigma\lto[]\lambda\actor[]p{\bsend qe;B_p}\parp\actor[]q{\brecv px;B_q}\parp N'',\sigma'$.
    The latter network extracts to $\gencom;C'''$, and Lemma~\ref{lem:extract-equiv} allows us to
    conclude that $\actor[]p{\bsend qe;B_p}\parp\actor[]q{\brecv px;B_q}\parp N''\rwto C''$.

    The case where $N,\sigma\lto[]\lambda N'',\sigma$ is similar, using Lemma~\ref{lem:sound-fin}.

  \item By structural induction on $C$.
    We detail one representative case.

    Suppose that $C$ is $\gencom;C'$.
    Then $\epp C$ can be written as $\actor[]p{\bsend qe;B_p}\parp\actor[]q{\brecv px;B_q}\parp N'$,
    where $\epp{C'}=\actor[]p{B_p}\parp\actor[]q{B_q}\parp N'$, and the thesis follows trivially by
    the induction hypothesis.\qedhere
  \end{enumerate}
\end{proof}
As we show later, the second part of this theorem does not hold in the presence of recursive
definitions.

The definition of $\rwto$ is convenient for finite SP: it is simple, and easy to analyse.
However, when we add the possibility of infinite behaviour, it will in general not be the case that
a network can be rewritten to a choreography in finitely many steps.
Therefore, we now restate extraction by means of constructing and analysing a particular graph.
This alternative method, which is the hallmark of our development, is easily seen to be equivalent
to the previous definition -- but it can be extended to the whole language of SP.

We start by introducing an abstract semantics for networks, $N \lto\alpha N'$, defined as in
Figure~\ref{fig:sp_semantics} with the following two differences: (i)~the state $\sigma$ is removed,
and (ii)~the rules for value communication and conditionals are replaced by those in
Figure~\ref{fig:asp_semantics}.
In particular, conditionals are nondeterministic in this semantics.

Labels $\alpha$ in the abstract semantics are like $\lambda$, but the labels for communications now
contain expressions and the variable for storing the result (see the new rule \rname{S}{Com}); in
all omitted rules, the label is the same as before.
We write $N\lmto{\til\alpha} N'$ for $N\lto{\alpha_1}\cdots\lto{\alpha_n}N'$.
\begin{figure}
  \begin{eqnarray*}
    &\infer[\rname{S}{Com}]{
      \actor[{\procs[p]}]p{\bsend qe;B_1} \parp
      \actor[{\procs[q]}]q{\brecv px;B_2}
      \lto[]{\com peqx}
      \actor[{\procs[p]}]p{B_1} \parp
      \actor[{\procs[q]}]q{B_2}
    }{}
    \\[1ex]
    &\infer[\rname{S}{Then}]{
      \actor[{\procs[p]}]p{\bcond e{B_1}{B_2}}
      \lto[]{\lthen pe}
      \actor[{\procs[p]}]p{B_1}
    }{}
    \\[1ex]
    &\infer[\rname{S}{Else}]{
      \actor[{\procs[p]}]p{\bcond e{B_1}{B_2}}
      \lto[]{\lelse pe}
      \actor[{\procs[p]}]p{B_2}
    }{}
  \end{eqnarray*}
  \caption{Abstract semantics for Stateful Processes. Besides these rules, the semantics includes all other rules in Figure~\ref{fig:sp_semantics} with the state removed.}
  \label{fig:asp_semantics}
\end{figure}

\begin{definition}
  \label{defn:extr-graph}
  Let $N$ be a network.
  The \emph{Abstract Execution Space (AES)} of $N$ is the directed graph obtained by considering all
  possible abstract reduction paths from $N$.
  Its vertices are all the networks $N'$ such that $N\lmto{\til\alpha} N'$, and there is an edge
  between two vertices $N_1$ and $N_2$ labelled $\alpha$ if $N_1\lto\alpha N_2$.
  
  A \emph{Symbolic Execution Graph (SEG)} for $N$ is a subgraph of its AES that contains $N$ and
  such that each vertex $N'\neq\nil$ has either one outgoing edge labelled by an interaction $\eta$ or two
  outgoing edges labelled $\lthen pe$ and $\lelse pe$, respectively.
\end{definition}

Intuitively, the AES of $N$ represents all possible evolutions of $N$ (each such evolution is a path
in this graph).
A SEG fixes the order of execution of actions, but still abstracts from the state (and thus
considers both branches of conditionals).
If $N$ is a network in finite SP, these graphs are trivially finite.

\begin{example}
  We revisit the networks in Example~\ref{ex:non-deterministic}.
  \begin{itemize}
  \item Network $N_1$ in the example has the following AES.
    \[\xymatrix{
      &
      \actor[] p{\bsend qe} \parp\actor[]q{\brecv px}
      \parp\actor[]r{\bsend s{e'}} \parp\actor[]s{\brecv ry}
      \ar[dl]_{\gencom} \ar[dr]^{\com r{e'}sy}
      \\
      \actor[]r{\bsend s{e'}} \parp\actor[]s{\brecv ry}
      \ar[dr]_{\com r{e'}sy}
      &&
      \actor[]p{\bsend qe} \parp\actor[]q{\brecv px}
      \ar[dl]^\gencom
      \\
      & \nil
    }\]
    
    This AES admits two SEGs, namely the two paths from the top node to the bottom node.
    Reading the labels of this path, one obtains the two choreographies that can be extracted from
    this network.

  \item Network $N_2$ illustrates how conditionals are treated.
    This network's AES, which coincides with its SEG, is the following.
    \[\xymatrix{
      & \smallnet{
        \actor[]p{\bcond e{\bsel q\lleft;\bsend q1}{\bsel q\lright;\brecv qx}}
        \parp
        \actor[]q{\bbranch p{\lleft:\brecv py,\ \lright:\bsend p2}}}
      \ar[]+D;[dl]+U_{\lthen pe} \ar'[]+D;[dr]+0^{\lelse pe} [dr]+0;[ddr]
      \\
      \smallnet{
        \actor[]p{\bsel q\lleft;\bsend q1} \parp
        \actor[]q{\bbranch p{\lleft:\brecv py,\ \lright:\bsend p2}}
      } \ar[dd]_{\sel pq\lleft}
      && \\
      && \smallnet{
        \actor[]p{\bsel q\lright;\brecv qx} \parp
        \actor[]q{\bbranch p{\lleft:\brecv py,\ \lright:\bsend p2}}
      } \ar[d]^{\sel pq\lright}
      \\
      \actor[]p{\bsend q1} \parp \actor[]q{\brecv py} \ar[]+D;[dr]_{\com p1qy}
      && \actor[]p{\brecv qx} \parp \actor[]q{\bsend p2} \ar[]+D;[dl]^{\com q2px}
      \\
      & \actor[]p\nil \parp \actor[]q\nil
    }\]
    Again, reading the labels on the edges of this graph, one obtains the choreography
    $\cond pe{\left(\sel pq\lleft;\com p1qy\right)}{\left(\sel pq\lright;\com q2px\right)}$, which
    describes the global behaviour of the original network.

  \item Finally, network $N_3$ gives the following AES.
    \[\xymatrix{
      & \smallnet{\actor[]p{\bsend q1;\bsend r2}
        \parp\actor[]q{\brecv px;\bsend r3}
        \parp\actor[]r{\bcond e{\brecv py}{\brecv qy}}}
      \ar[]+D;[dl]+U_{\lthen re} \ar[]+D;[dd]^{\com p1qx} \ar[]+D;[dr]+U^{\lelse re} \\
      \smallnet{\actor[]p{\bsend q1;\bsend r2}
        \parp\actor[]q{\brecv px;\bsend r3}
        \parp\actor[]r{\brecv py}} \ar[]+D;[dd]+U_{\com p1qx}
      &&
      \smallnet{\actor[]p{\bsend q1;\bsend r2}
        \parp\actor[]q{\brecv px;\bsend r3}
        \parp\actor[]r{\brecv qy}} \ar[]+D;[dd]+U^{\com p1qx} \\
      &
      \smallnet[6em]{\actor[]p{\bsend r2}\parp\actor[]q{\bsend r3}
        \parp\actor[]r{\bcond e{\brecv py}{\brecv qy}}}
      \ar[]+D;[dl]_{\lthen re}  \ar[]+D;[dr]^{\lelse re}
      \\
      \actor[]p{\bsend r2}\parp\actor[]q{\bsend r3}\parp\actor[]r{\brecv py} \ar[]+D;[d]^{\com p2ry}
      &&
      \actor[]p{\bsend r2}\parp\actor[]q{\bsend r3}\parp\actor[]r{\brecv qy} \ar[]+D;[d]^{\com q3ry}
      \\
      \actor[]p\nil\parp\actor[]q{\bsend r3}\parp\actor[]r\nil
      &&
      \actor[]p{\bsend r2}\parp\actor[]q\nil\parp\actor[]r\nil
    }\]
    There are two SEGs for this AES:
    \[\xymatrix{
      & \smallnet{\actor[]p{\bsend q1;\bsend r2}
        \parp\actor[]q{\brecv px;\bsend r3}
        \parp\actor[]r{\bcond e{\brecv py}{\brecv qy}}}
      \ar[]+D;[dl]+U_{\lthen re} \ar[]+D;[dr]+U^{\lelse re} \\
      \smallnet{\actor[]p{\bsend q1;\bsend r2}
        \parp\actor[]q{\brecv px;\bsend r3}
        \parp\actor[]r{\brecv py}} \ar[]+D;[d]+U_{\com p1qx}
      &&
      \smallnet{\actor[]p{\bsend q1;\bsend r2}
        \parp\actor[]q{\brecv px;\bsend r3}
        \parp\actor[]r{\brecv qy}} \ar[]+D;[d]+U^{\com p1qx} \\
      \actor[]p{\bsend r2}\parp\actor[]q{\bsend r3}\parp\actor[]r{\brecv py} \ar[]+D;[d]^{\com p2ry}
      &&
      \actor[]p{\bsend r2}\parp\actor[]q{\bsend r3}\parp\actor[]r{\brecv qy} \ar[]+D;[d]^{\com q3ry}
      \\
      \actor[]p\nil\parp\actor[]q{\bsend r3}\parp\actor[]r\nil
      &&
      \actor[]p{\bsend r2}\parp\actor[]q\nil\parp\actor[]r\nil
    }\]
    and
    \[\xymatrix{
      & \smallnet{\actor[]p{\bsend q1;\bsend r2}
        \parp\actor[]q{\brecv px;\bsend r3}
        \parp\actor[]r{\bcond e{\brecv py}{\brecv qy}}}
      \ar[]+D;[d]^{\com p1qx} \\
      &
      \smallnet{\actor[]p{\bsend r2}\parp\actor[]q{\bsend r3}
        \parp\actor[]r{\bcond e{\brecv py}{\brecv qy}}}
      \ar[]+D;[dl]^{\lthen re}  \ar[]+D;[dr]_{\lelse re}
      \\
      \actor[]p{\bsend r2}\parp\actor[]q{\bsend r3}\parp\actor[]r{\brecv py} \ar[]+D;[d]^{\com p2ry}
      &&
      \actor[]p{\bsend r2}\parp\actor[]q{\bsend r3}\parp\actor[]r{\brecv qy} \ar[]+D;[d]^{\com q3ry}
      \\
      \actor[]p\nil\parp\actor[]q{\bsend r3}\parp\actor[]r\nil
      &&
      \actor[]p{\bsend r2}\parp\actor[]q\nil\parp\actor[]r\nil
    }\]
    Both SEGs end in deadlocked networks, in line with the fact that $N_3$ cannot be extracted to a
    choreography.
    Representing these networks by $\dlock$ and reading the labels on the edges in these graphs
    allows us to reconstruct the extracted extended choreographies
    $\com p1qx;\cond re{\left(\com p2ry;\dlock\right)}{\left(\com q3ry;\dlock\right)}$ and
    $\cond re{\left(\com p1qx;\com p2ry;\dlock\right)}{\left(\com p1qx;\com q3ry;\dlock\right)}$.
    \eoe
  \end{itemize}
\end{example}

As these examples illustrate, there is a strong connection between these graphs and the previous
definition of extraction: each rule in Definition~\ref{defn:extraction} naturally corresponds to an
edge, except for the first (which characterises terminated networks) and the last (which
characterises deadlocked networks). Therefore,
each particular sequence of steps extracting a choreography corresponds to a SEG, and
conversely.

\begin{lemma}
  The AES for any network in finite SP is a directed acyclic graph (DAG).
\end{lemma}
\begin{proof}
  Since there are no procedure calls, every reduction strictly decreases the size of the network
  (measured by the number of nodes in its abstract syntax tree).
  Therefore no network can ever reduce to itself in any number of steps, and as such no AES can
  have loops.
\end{proof}
As a consequence, every SEG for a network in finite SP is also a DAG.

\begin{definition}
  Let $S$ be a SEG for a network.
  The extended choreography body extracted from node $N$, $\extract[S]N$, is defined inductively as
  follows.
  \begin{itemize}
  \item $\extract[S]\nil=\nil$
  \item If $N$ has no descendants and $N\neq\nil$, then $\extract[S]N=\dlock$.
  \item If $N$ has one descendant $N'$ and the edge from $N$ to $N'$ has label $\eta$, then
    $\extract[S]N=\eta;\extract[S]N'$.
  \item If $N$ has two descendants $N'$ and $N''$ and the edges from $N$ to those nodes are labelled
    $\lthen pe$ and $\lelse pe$, respectively, then
    $\extract[S]N=\cond pe{\extract[S]N'}{\extract[S]N''}$.
  \end{itemize}
\end{definition}

\begin{example}
  The (extended) choreographies informally presented in the previous example correspond exactly to
  the (extended) choreographies extracted from the given SEGs.
  \eoe
\end{example}

As the examples suggest, this new notion of extraction coincides precisely with the old one.
\begin{lemma}
  Let $N$ be a network.
  \begin{enumerate}[(i)]
  \item If $S$ is a SEG for $N$, then $\extract N\rwtot\extract[S]N$.
  \item For every choreography body $C$, if $\extract N\rwtot C$, then there exists a SEG $S$ for
    $N$ such that $\extract[S]N=C$.
  \end{enumerate}
\end{lemma}
\begin{proof}\mbox{}
  \begin{enumerate}[(i)]
  \item Straightforward by induction on the definition of $\extract[S]N$, since every case in its
    definition corresponds directly to a rule in the definition of $\rwto$.
  \item The sequence of reductions in $\extract N\rwtot C$ defines a graph $S$ as follows:
    \begin{itemize}
    \item an application of the first or last rule does not add anything to the graph;
    \item an application of the second or third rule generates an edge from the network on the left
      to the reductum network on the right, labelled with the choreography action that is introduced
      by the rule;
    \item an application of the fourth rule generates two edges from the network on the left to each
      reductum network on the right, labelled by the appropriate conditional label.
    \end{itemize}
    It is immediate to check that $S$ is a SEG for $N$, and that $\extract[S]N=C$.\qedhere
  \end{enumerate}
\end{proof}

\subsection{Adding recursion}
\label{sec:recursion}

Formulating extraction in terms of SEGs allows us to extend it to networks with recursive
definitions.
The tricky step is defining the AES: abstract executions of a network can be infinite, and due to
recursion unfolding there are in general infinite possible future states of a network with truly
recursive definitions.
Defining extraction from such infinite graphs would be problematic already, since choreographies are
finite; furthermore, we are interested in \emph{computing} extracted choreographies, which requires
at least building a SEG.

To ensure finiteness, we restrict the applications of rule~\rname{S}{Unfold} in the abstract
semantics (Figure~\ref{fig:asp_semantics}).
\begin{enumerate}[(i)]
\item Rule \rname{S}{Unfold} can only be applied inside a derivation occurring in the first premise
  of rule~\rname{S}{Struct}.
\item If rule~\rname{S}{Unfold} is applied to process $\pid p$ inside a derivation proving
  $N\precongr[] M$, then $N(\pid p)$ is a procedure call.
\item If rule~\rname{S}{Unfold} is applied to process $p$ inside a derivation using rule
  \rname{S}{Struct}, then process $\pid p$ appears in the label $\lambda$ of the reduction.
\end{enumerate}
In other words: we only allow unfolding recursive definitions in order to execute a reduction that
would otherwise not be enabled.

With these restrictions, the AES and SEGs for a network are defined as in the finite case.
However, these graphs no longer need to be DAGs, since a network may evolve into itself after some
reductions.

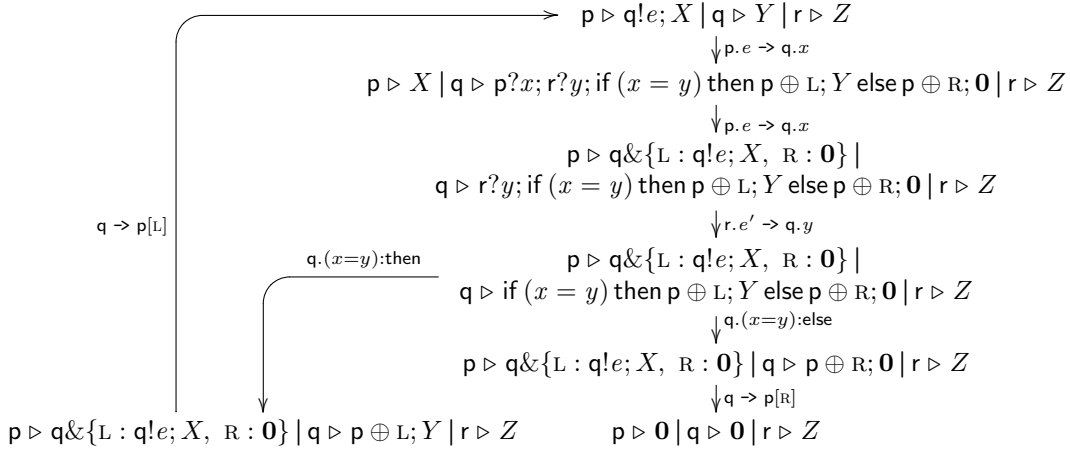
\begin{figure}[t]
  \[\xymatrix@R=1em{
    & \makebox[0cm][c]{$\actor p{\bsend qe;X} \parp \actor qY \parp \actor rZ$}
    \ar[d]^{\com peqx}
    \\
    & \makebox[0cm][c]{$\actor pX \parp \actor q{\brecv px;\brecv ry;\bcond{(x=y)}{\bsel p\lleft;Y}{\bsel p\lright;\nil}} \parp \actor rZ$}
    \ar[d]^{\com peqx}
    \\
    & {\begin{array}c \displaystyle\actor p{\bbranch q{\lleft:\bsend qe;X,\ \lright:\nil}} \parp {} \\
        \displaystyle \actor q{\brecv ry;\bcond{(x=y)}{\bsel p\lleft;Y}{\bsel p\lright;\nil}} \parp \actor rZ
        \end{array}}
    \ar[d]^{\com r{e'}qy}
    \\
    & {\begin{array}c \displaystyle\actor p{\bbranch q{\lleft:\bsend qe;X,\ \lright:\nil}} \parp {} \\
        \displaystyle \actor q{\bcond{(x=y)}{\bsel p\lleft;Y}{\bsel p\lright;\nil}} \parp \actor rZ
        \end{array}}
    \ar '[]+L `l[ddl]_{\lthen q{(x=y)}} [ddl]
    \ar[d]^{\lelse q{(x=y)}}
    \\
    & \makebox[8em][c]{$\actor p{\bbranch q{\lleft:\bsend qe;X,\ \lright:\nil}} \parp \actor q{\bsel p\lright;\nil} \parp \actor rZ$}
    \ar[d]^{\sel qp\lright}
    \\
    \makebox[6em][c]{$\actor p{\bbranch q{\lleft:\bsend qe;X,\ \lright:\nil}} \parp \actor q{\bsel p\lleft;Y} \parp \actor rZ$}
    \ar '[]+UL `u[uuuuur]-<6em,0em>^{\sel qp\lleft} [uuuuur]-<6em,0em>
    & \makebox[0em][c]{$\actor p\nil \parp \actor q\nil \parp \actor rZ$}
  }\]
  \caption{The AES and SEG for the network in Example~\ref{ex:aes}.}
  \label{fig:aes}
\end{figure}

\begin{example}
  \label{ex:aes}
  Consider the network
  \[
  \actor p{\bsend qe;X}
  \parp
  \actor qY
  \parp
  \actor rZ
  \]
  where procedures $X$, $Y$, and $Z$ are defined at $\pid p$, $\pid q$, and $\pid r$, respectively, as
  \begin{align*}
    X &= \bsend qe;\bbranch q{\lleft:\bsend qe;X,\ \lright:\nil} \\
    Y &= \brecv px;\brecv px;\brecv ry; \bcond{(x=y)}{\bsel p\lleft;Y}{\bsel p\lright;\nil} \\
    Z &= \bsend q{e'};Z
  \end{align*}

  This network generates the AES in Figure~\ref{fig:aes}.
  Since execution of this network is deterministic, the same graph is also its SEG.
  \eoe
\end{example}

The key insight to define extraction in this case is that the definitions of recursive procedures
are extracted from the loops in the SEG, rather than from the recursive definitions in the source
network.

\begin{definition}
  \label{def:DAG-ification}
  Let $S$ be a SEG for a network $N$.
  A \emph{loop node} is a node $n$ in $S$ such that: (i)~$n$ has more than one incoming edge or
  (ii)~$n$ is the initial node labelled $N$ and $n$ has at least one incoming edge.
  
  The \emph{DAG-ification} of $S$ is the graph $S^D$ defined as follows.
  \begin{itemize}
  \item The nodes of $S^D$ are all the nodes of $S$ together with new nodes $X_n$ for each loop node
    $n$.
  \item For each edge in $S$ from $n$ to $n'$, $S^D$ contains one edge with source $n$ and target $X_{n'}$,
    if $n'$ is a loop node, and source $n$ and target $n'$, otherwise.
  \end{itemize}
\end{definition}

\begin{lemma}
  Graph $S^D$ is a DAG.
\end{lemma}
\begin{proof}
  Suppose there is a cycle $n_1,n_2\ldots,n_k=n_1$ in $S$.
  If one of $n_1,\ldots,n_k$ is the initial node, then this path is no longer a path in $S^D$ by
  construction.
  Otherwise, one of these nodes must have at least two incoming edges (since all nodes are
  accessible from the initial node), which again implies that it is no longer a path in $S^D$.
\end{proof}

From the root node of each connected component of $S^D$, we can extract a choreography as before,
adding the rule $\extract[S^D]{X_n}=X_n$ where $X_n$ is a procedure name.
\begin{definition}
  \label{defn:extr-unsound}
  The choreography extracted from $N$, $\extract[S]N$ is defined as follows.
  \begin{itemize}
  \item The set of procedure definitions is
    $\{\rec{X_{n'}}{\extract[S^D]{n'}\mid n'\mbox{ is a loop node in $S$}}\}$.
  \item The main choreography is $X_n$, if the starting node $n$ is a loop node, and
    $\extract[S^D]N$, otherwise.
  \end{itemize}
\end{definition}

\begin{example}
  Consider the SEG in Figure~\ref{ex:aes}.
  To extract a choreography, we split the topmost node into two nodes; the new node is labelled with
  a procedure identifier $X$, which is the target of the upgoing arrow in the figure.
  Thus, $X$ is extracted to
  \[ \com peqx;\com peqx;\com r{e'}qy;\cond q{(x=y)}{\sel qp\lleft;X}{\sel qp\lright;X} \]
  and the extracted choreography itself is simply $X$.

  The body of $X$ is not projectable (the branches for $\pid r$ are not mergeable,
  cf.~\cite{CM20}), but it faithfully describes the behaviour of the original network.
  \eoe
\end{example}

The procedure in Definition~\ref{defn:extr-unsound} always terminates, but sometimes
it extracts incomplete choreographies that lack some behaviours from the original network.
We illustrate the possible problems with some examples.

\begin{example}
  \label{ex:starve}
  Consider the network $N$ defined as $\actor pX \parp \actor qY \parp \actor rZ \parp \actor sW$,
  where the recursive procedures $X$, $Y$, $Z$, and $W$ are as follows.
  \[
  \rec X{\bsend qe;X} \qquad
  \rec Y{\brecv px;Y} \qquad
  \rec Z{\bsend s{e'};Z} \qquad
  \rec W{\brecv ry;W}
  \]
  The AES for $N$ is:
  \[\xymatrix{
   \actor pX\parp\actor qY\parp\actor rZ\parp\actor sW
   \ar@(d,l)[]+L^{\com peqx} \ar@(d,r)[]+R_{\com r{e'}sy}
  }\]

  There are two SEGs for this AES:
  \[\xymatrix{
    \actor pX\parp\actor qY\parp\actor rZ\parp\actor sW
    \ar@(d,l)[]+L^{\com peqx}
    & \text{ and } &
    \actor pX\parp\actor qY\parp\actor rZ\parp\actor sW
    \ar@(d,r)[]+R_{\com r{e'}sy}
  }\]
  which extract to choreographies consisting of a call to procedure $X$, defined as
  $\rec X{\com peqx;X}$ and $\rec X{\com r{e'}sy;X}$, respectively, none of which captures all the behaviours of~$N$.
  \eoe
\end{example}

\begin{example}
  \label{ex:finite}
  A similar situation may occur if there are processes with finite behaviour (no procedure calls):
  the network $\actor pX \parp \actor qY \parp \actor r{\bsend s{e'}} \parp \actor s{\brecv ry}$
  where $\rec X{\bsend qe;X}$ and $\rec Y{\brecv px;Y}$ can be extracted to the choreography $Z$,
  with $\rec Z{\com peqx};Z$, where $\pid r$ and $\pid s$ never communicate.
  \eoe
\end{example}

Both these examples exhibit a form of starvation: there is a loop involving some processes that can
reduce (as can be seen in the AES), but they are not allowed to do so in a particular SEG.
Example~\ref{ex:starve} is particularly relevant, since there is no SEG where all involved processes
reduce.

In order to avoid such situations, we change the definitions of AES and SEG slightly.
We annotate all processes in networks with either $\circ$ (unmarked) or $\bullet$ (marked).
In the initial network, all processes are unmarked.
Processes are marked when they are involved in a reduction; the marking is reset when all processes
are marked.

To make this formal, we extend the semantics to annotated networks as follows.
Let $N$ and $N'$ be annotated networks, and $N^-$ and ${N'}^-$ be the underlying networks obtained
by erasing the annotations.
Then $N\lto\alpha N'$ if:
\begin{itemize}
\item $N^-\lto\alpha{N'}^-$;
\item all processes in $N'$ are unmarked iff all unmarked processes in $N$ appear in $\alpha$;
\item otherwise, a process is marked in $N'$ iff it is marked in $N$ or it appears in $\alpha$.
\end{itemize}

\begin{definition}
  \label{defn:valid-seg}
  A SEG for a network $N$ is \emph{valid} if all its loops include a node where all processes are
  unmarked.

  A network $N$ extracts to a choreography $C$ if $C$ can be constructed (as in
  Definition~\ref{defn:extr-unsound}) from a valid SEG for $N$.
\end{definition}
In a valid SEG, every process is guaranteed to reduce at least once inside every loop.

\begin{example}
  The AES for the annotated network $N$ in Example~\ref{ex:starve} is:
  \[\xymatrix@R=1em{
    &\makebox[2em][c]{$\actor{p^\circ}X\parp\actor{q^\circ}Y\parp\actor{r^\circ}Z\parp\actor{s^\circ}W$}
    \ar@/^/[dl]_(.4){\com peqx}
    \ar@/_/[dr]^(.4){\com r{e'}sy}
    \\
    \makebox[10em][c]{$\actor{p^\bullet}X\parp\actor{q^\bullet}Y\parp\actor{r^\circ}Z\parp\actor{s^\circ}W$}
    \ar@/^/[ur]+DL+<-5.7em,.7em>^{\com r{e'}sy}
    \ar@(dr,dl)[]+DL_{\com peqx}
    &&\makebox[10em][c]{$\actor{p^\circ}X\parp\actor{q^\circ}Y\parp\actor{r^\bullet}Z\parp\actor{s^\bullet}W$}
    \ar@/_/[ul]+DR+<5.7em,.7em>_{\com peqx}
    \ar@(dl,dr)[]+DR^{\com r{e'}sy}
  }\]
  This AES now has the following two SEGs:
  \[\xymatrix@R=1em{
    \actor{p^\circ}X\parp\actor{q^\circ}Y\parp\actor{r^\circ}Z\parp\actor{s^\circ}W
    \ar@/^/[d]^{\com peqx}
    &&
    \actor{p^\circ}X\parp\actor{q^\circ}Y\parp\actor{r^\circ}Z\parp\actor{s^\circ}W
    \ar@/^/[d]^{\com r{e'}sy}
    \\
    \actor{p^\bullet}X\parp\actor{q^\bullet}Y\parp\actor{r^\circ}Z\parp\actor{s^\circ}W
    \ar@/^/[u]^{\com r{e'}sy}
    &&
    \actor{p^\circ}X\parp\actor{q^\circ}Y\parp\actor{r^\bullet}Z\parp\actor{s^\bullet}W
    \ar@/^/[u]^{\com peqx}
  }\]
  Observe that the self-loops from the AES are discarded because they do not go through a node where
  all processes are unmarked.

  From these SEGs, we can extract two definitions for $X$:
  \[
  \rec{X}{\com peqx;\com r{e'}sy;X}
  \qquad\mbox{ and }\qquad
  \rec{X}{\com r{e'}sy;\com peqx;X}
  \]
  and both of these definitions correctly capture all behaviours of the network.
  \eoe
\end{example}

Validity implies, however, that there are some non-deadlocked networks that are not extractable,
such as $\actor pX\parp \actor qY\parp \actor rZ$ where $\rec X{\bsend qe;X}$, $\rec Y{\brecv px;Y}$
and $\rec Z{\brecv py;Z}$, for which there is no valid SEG.
This is to be expected, since deadlock-freedom is undecidable in SP.

In practice, there are situations where livelocks are acceptable, namely in the presence of a service that is designed to be used only when necessary. In Section~\ref{sec:impl-services} we briefly discuss how to deal with such cases.




\subsection{Soundness and completeness}

Since extraction ignores the definition of procedures, it is simple to find counterexamples to the
second part of Theorem~\ref{thm:correct}.

\begin{example}
  Consider the very simple choreography \[\chorterm{X:=\gencom;\gencom;X}{\gencom;X}\,.\]
  Its projection is the network
  \[
  \actor p{\procterm{X:=\bsend qe;\bsend qe;X}{\bsend qe;X}}
  \parp
  \actor q{\procterm{X:=\brecv px;\brecv px;X}{\brecv px;X}}
  \]
  which extracts to the choreography $\chorterm{X:=\gencom;\gencom;X}X$.
  \eoe
\end{example}

We show that the first part of this result still holds, from which it follows that the analogue of
Lemma~\ref{lem:sound-fin} also applies.
This proof is divided into several steps.

Throughout this section, let $N$ be a network, $\chorterm\procs C$ be the choreography
extracted from $N$ for a particular SEG $\mathcal G$, and $\sigma$ be a state.
We consider the (possibly infinite) sequences $\{\lambda_i\}_{i\in I}$, $\{C_i\}_{i\in I}$, and
$\{\sigma_i\}_{i\in I}$ defined as:
\begin{itemize}
\item $C_0=C$;
\item $\sigma_0=\sigma$;
\item for each $i$, $\lambda_i$ is the label of the reduction executing the head action in $C_i$
  (the only action that can be executed without applying any of the structural congruence rules
  other than~\rname{C}{Unfold});
\item for each $i$, $C_{i+1}$ and $\sigma_{i+1}$ are the only choreography and state such that
  $C_i,\sigma_i\lto{\lambda_i}C_{i+1},\sigma_{i+1}$;
\item $I=\{0,\ldots,n\}$ if $C_n$ is $\nil$ for some $n$, and $\mathbb N$ otherwise.
\end{itemize}
Observe that $C_{i+1}$ and $\sigma_{i+1}$ are well-defined, since the semantics of CC completely
determines these terms given $C_i$, $\sigma_i$, and $\lambda_i$.

\begin{lemma}
  There exists a sequence $\{n_i\}_{i\in I}$ in $\mathcal G$ such that there is an edge
  $n_i\lto[]{\lambda'_i}n_{i+1}$, where $\lambda'_i$ is the label corresponding to $\lambda_i$ in
  the abstract semantics for SP.
\end{lemma}
\begin{proof}
  By induction on $i$.
  We take $n_0$ to be the starting node in the construction of $\mathcal G$.
  By construction of $C$, for each $i$, there must be an outgoing edge labelled with $\lambda'_i$,
  and we define $n_{i+1}$ as the target of that edge.
\end{proof}

\begin{lemma}
  There exists a sequence of networks $\{N_i\}_{i\in I}$ such that $N_0=N$ and
  $N_i,\sigma_i\lto[]{\lambda_i}N_{i+1},\sigma_{i+1}$.
\end{lemma}
\begin{proof}
  We prove by induction that $N_0,\sigma_0\lmto{\lambda_0\ldots\lambda_{i-1}}N_i,\sigma_i$, where
  $N_i$ is the network labelling the node $n_i$ defined in the previous lemma.
  This trivially holds for $i=0$.
  Now assume by induction hypothesis that it holds for $i-1$.
  Given how SEGs are constructed, $\lambda'_i$ is an abstraction of an action that $N_{i-1}$ can
  execute, and the only possible action corresponding to it is $\lambda_i$ (since the details
  missing in the abstraction are uniquely defined by $\sigma_{i-1}$, and they coincide for
  choreographies and networks).
  The semantics of SP guarantees that there exist unique $N'$ and $\sigma'$ such that
  $N_{i-1},\sigma_{i-1}\lto[]{\lambda_i}N',\sigma'$.
  Since the abstract and concrete semantics act in the same way on networks, $N'=N_i$; and since
  an inspection of the rules for the semantics of CC and SP establishes that $\sigma'=\sigma_i$.
\end{proof}

\begin{lemma}
  \label{lem:reduces-iff}
  For every $i\in I$ and reduction label $\lambda$, $C_i,\sigma_i$ can execute a reduction labelled
  by $\lambda$ iff $N_i,\sigma_i$ can execute a reduction labelled by $\lambda$.
\end{lemma}
\begin{proof}
  Assume that $C_i,\sigma_0\lto\lambda C',\sigma'$ for some $C'$ and $\sigma'$.
  Let $j\geq i$ be the minimal index such that $\lambda$ and $\lambda_j$ share process names.
  Since structural precongruence can only exchange actions that do not share process names and the
  semantics of CC only allows one action for each process at each point, it immediately follows that
  $\lambda=\lambda_j$.
  Furthermore, since no action in $\lambda_i,\ldots,\lambda_{j-1}$ shares process names with
  $\lambda$, it follows that the behaviour of the processes involved in $\lambda$ is unchanged in
  $N_i,\ldots,N_j$ and that $\sigma_i(\pid p)=\sigma_j(\pid p)$ for every such process $\pid p$.
  Since $N_j,\sigma_j$ can execute $\lambda$ and the conditions for executing an action are local to
  the processes involved in that action, this implies that $N_i,\sigma_i\lto[]\lambda N',\sigma'$
  for some network $N'$ -- the same argument as in previous proofs implies that the resulting state
  must be $\sigma'$.

  Now assume that $N_i,\sigma_i\lto[]\lambda N',\sigma'$ for some $N'$ and $\sigma'$.
  Since the processes involved in $\lambda$ cannot participate in any other reductions, $\lambda$
  is enabled in all nodes of $\mathcal G$ until an edge labelled by its abstract counterpart is
  traversed -- in other words, $N_i,\ldots,N_j$ can all execute $\lambda$ for the least $j\geq i$
  such that $\lambda=\lambda_j$.
  Furthermore, such a $j$ must exist due to the fairness conditions imposed by
  Definition~\ref{defn:valid-seg}: since $\mathcal G$ is a valid SEG, either execution of $N_i$
  terminates (in which case $\lambda$ must have been executed) or there is a loop in the SEG, and
  every process in the network must reduce at least once inside that loop (and again $\lambda$ must
  be executed in that loop).
  Since $\lambda$ shares no process names with any actions in $\lambda_i,\ldots,\lambda_{j-1}$, it
  also follows that $C_i$ can execute it, and as before the resulting state must be $\sigma'$.
  We thus conclude that $C_i,\sigma_i\lto\lambda C',\sigma'$ for some $C'$.
\end{proof}

The next lemma is a property of CC not directly related to extraction.
\begin{lemma}
  \label{lem:chor-perm}
  Let $\til\lambda'=\lambda'_0,\ldots,\lambda'_j$ be a (finite) sequence of reduction labels such
  that $C,\sigma\lcmto{\til\lambda'}C',\sigma'$.
  Then there exist $n\in\mathbb N$ and a permutation $\pi:\{0,\ldots,n\}\to\{0,\ldots,n\}$ such that
  $\lambda'_i=\lambda(\pi(i))$ for $i=0,\ldots,j$.
  Furthermore, $\pi$ can be obtained by repeatedly transposing consecutive pairs of labels that
  share no process names.
\end{lemma}
\begin{proof}[Proof (sketch).]
  This result is a corollary of the proof of confluence of CC from~\cite{CMP21a}, although it
  has not been stated in this form before.
  Confluence is proved by first showing that executing two independent actions in any order always
  yields the same result.
  This is extended by induction to sequences of actions, where the inductive case is split
  according to whether both sequences start with the same action.

  The current lemma follows from observing that we can choose a large enough $n$ such that all
  actions in $\til\lambda'$ occur in $\lambda_0,\ldots,\lambda_n$.
  Unfolding the proof of confluence as described above iteratively applies a transposition of
  consecutive independent actions to $\lambda_0,\ldots,\lambda_n$, until this sequence starts with
  $\lambda'_0,\ldots,\lambda'_j$.
  The composition of these transpositions yields the permutation $\pi$.
\end{proof}

\begin{lemma}
  \label{lem:net-perm}
  Let $\til\lambda'=\lambda'_0,\ldots,\lambda'_j$ be a (finite) sequence of reduction labels such
  that $N,\sigma\lmto{\til\lambda'}N',\sigma'$.
  Then there exist $n\in\mathbb N$ and a permutation $\pi:\{0,\ldots,n\}\to\{0,\ldots,n\}$ such that
  $\lambda'_i=\lambda(\pi(i))$ for $i=0,\ldots,j$.
  Furthermore, $\pi$ can be obtained by repeatedly transposing consecutive pairs of labels that
  share no process names.
\end{lemma}

\begin{lemma}
  \label{lem:perm-red}
  Let $\til\lambda'=\lambda'_0,\ldots,\lambda'_j$ be a prefix of any sequence of reduction labels
  obtained by repeatedly transposing consecutive elements of $\lambda$ that share no process names.
  Then there exist a choreography $C'$, a network $N'$ and a state $\sigma'$ such that
  $C,\sigma\lto{\til\lambda'}C',\sigma'$ and $N,\sigma\lto[]{\til\lambda'}N',\sigma'$.
  Furthermore, the actions that $C'$ and $N'$ can execute coincide.
\end{lemma}
\begin{proof}
  By induction on the number of transpositions applied.
  If this number is~$0$, then this is simply Lemma~\ref{lem:reduces-iff}.

  Assume by induction hypothesis that the thesis holds for $\{\lambda'_i\}_{i\in I}$ obtained by
  applying $n$ transpositions to consecutive actions in $\lambda$, and suppose that $\lambda'_i$ and
  $\lambda'_{i+1}$ share no process names.
  Note that the thesis holds for the sequence obtained by swapping these two labels for any
  $j\neq i$: for $j<i$ the sequence $\lambda'$ is unchanged, while confluence ensures that the
  result of executing $\lambda'_0,\ldots,\lambda'_i,\lambda'_{i+1}$ coincides with the result of
  executing $\lambda'_0,\ldots,\lambda'_{i+1},\lambda'_i$ for both $C$ and $N$.
  But as observed before, a reduction does not change the possible actions of processes not involved in it.
  Since $\lambda_i$ and $\lambda_{i+1}$ do not share any such processes, if
  $C,\sigma\lto{\lambda'_0,\ldots,\lambda'_{i-1},\lambda'_{i+1}}C'',\sigma''$, then the executable
  actions in $C''$ are those that were already available in the previous step, together with any
  actions unblocked by $\lambda_{i+1}$.
  Furthermore, the latter actions remain unchanged after executing $\lambda_i$.
  A similar reasoning applies to the executable actions in $N''$, where
  $N,\sigma\lto[]{\lambda'_0,\ldots,\lambda'_{i-1},\lambda'_{i+1}}N'',\sigma''$.
  Since the set of executable actions before and after executing $\lambda_{i+1}$ and $\lambda_i$
  coincide, the actions executable by $C''$ and $N''$ are defined in the same way, and therefore
  must also coincide.
\end{proof}

\begin{theorem}
  \label{thm:oc-ac-ap}
  If $C$ is a choreography extracted from a network $N$, then $N\bisim C$.
\end{theorem}
\begin{proof}
  Let $N$ be a network, $C$ be a choreography extracted from $N$, and $\sigma$ be a state.
  Define a relation $\mathcal R\subseteq\mathcal C\times\mathcal N$, where
  $\mathcal C=\{C'\mid C,\sigma\to^\ast C'\sigma'\mbox{ for some $\sigma'$}\}$ and
  $\mathcal N=\{N'\mid N,\sigma\to^\ast N',\sigma'\mbox{ for some $\sigma'$}\}$, as follows:
  $C'\mathcal R N'$ if $C,\sigma\lcmto{\til\lambda'}C',\sigma'$ and
  $N,\sigma\lmto{\til\lambda'}N',\sigma'$ for some sequence of actions $\til\lambda'$.

  We show that $\mathcal R$ is a bisimulation.
  Assume that $C'\mathcal R N'$.
  Then there exists a sequence of actions $\til\lambda'$ such that
  $C,\sigma\lcmto{\til\lambda'}C',\sigma'$ and $N,\sigma\lmto{\til\lambda'}N',\sigma'$.
  By Lemma~\ref{lem:chor-perm}, $\til\lambda'$ can be obtained from $\til\lambda$ by repeatedly
  permuting two consecutive independent actions and taking an initial segment of the result.
  By Lemma~\ref{lem:perm-red}, the actions that $C'$ and $N'$ can execute are therefore the same.
  For each such action $\alpha$, we can again apply Lemmas~\ref{lem:chor-perm}
  and~\ref{lem:perm-red} to the sequence $\til\lambda';\alpha$ to conclude that, if
  $C',\sigma'\lto\alpha C'',\sigma''$, then there exists $N''$ such that
  $N',\sigma'\lto[]\alpha N'',\sigma''$; conversely, if $N',\sigma'\lto[]\alpha N'',\sigma''$, then
  applying Lemmas~\ref{lem:net-perm} and~\ref{lem:perm-red} yields that
  $C',\sigma'\lto\alpha C'',\sigma''$ for some $C''$.
\end{proof}

\section{Implementation}
\label{sec:impl}
We now describe an implementation of the algorithm presented in
Section~\ref{sec:extraction}, with emphasis on the interesting
technical details.  The main challenge is computing a valid SEG for
the input network efficiently, or determining in reasonable time that
none exists; we follow the idea, given previously, of lazily expanding
the relevant parts of the AES until a valid SEG is found or we can
safely conclude that none exists.

\subsection{Overview}
\begin{figure}
  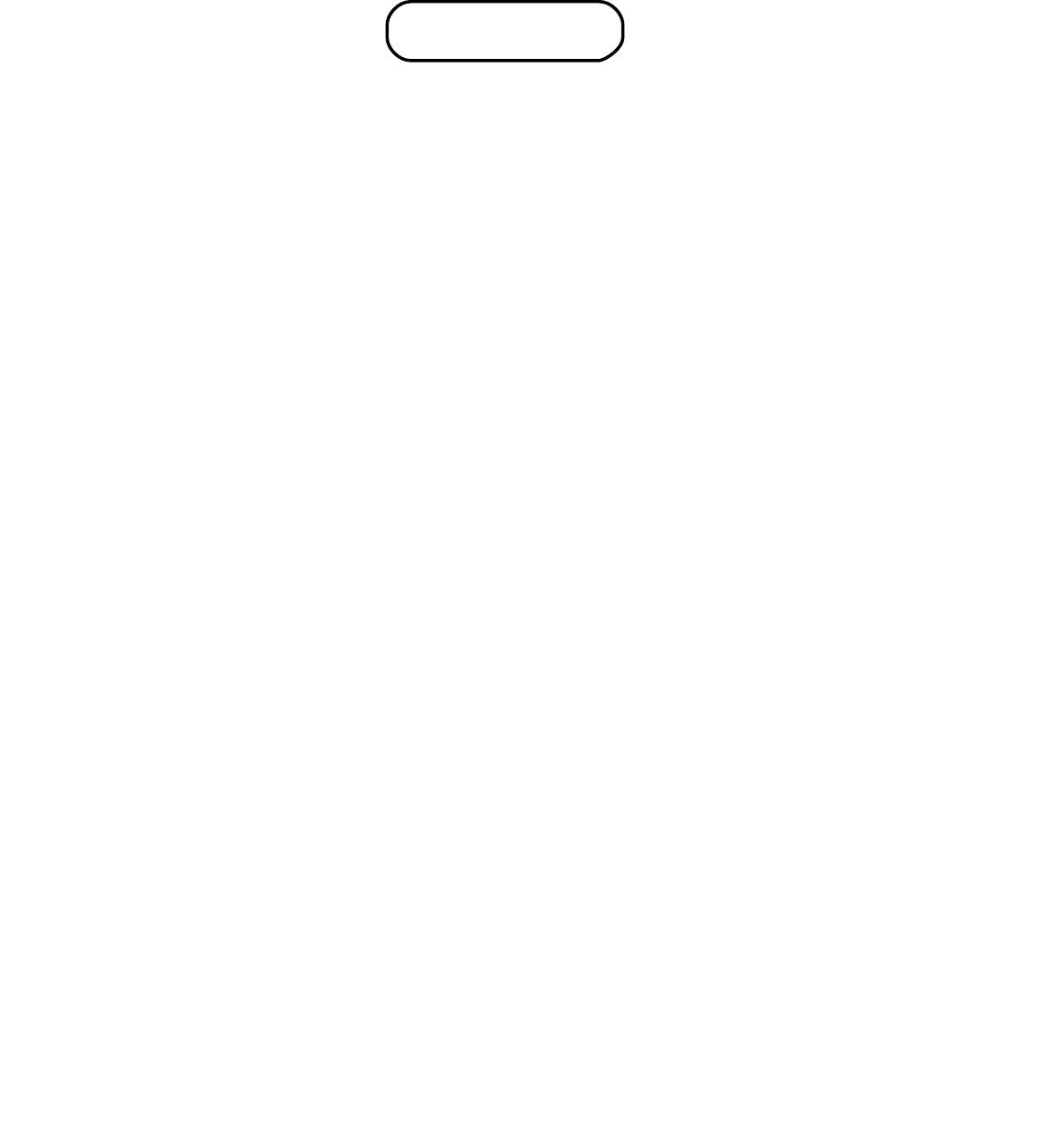
  \caption{Graphical depiction of \code{buildGraph}.}
  \label{fig:buildGraph}
\end{figure}

The extraction algorithm is implemented in a depth-first manner, starting with a single node (the
initial network, properly annotated), on which we call a method, \code{buildGraph}, graphically
described in Figure~\ref{fig:buildGraph}.
This method builds a list of all actions that the network can execute: a communication (of either a
value or a label) between two processes, or the execution of a conditional at a process.
This list includes actions that require unfolding procedure calls.
Then, the method tries to complete the SEG assuming that the first action in the list is executed,
returning \code{true} if this succeeds.
If this step fails, the next action in the list is considered.
If no action leads to success, \code{buildGraph} returns \code{false}.

\begin{figure}
  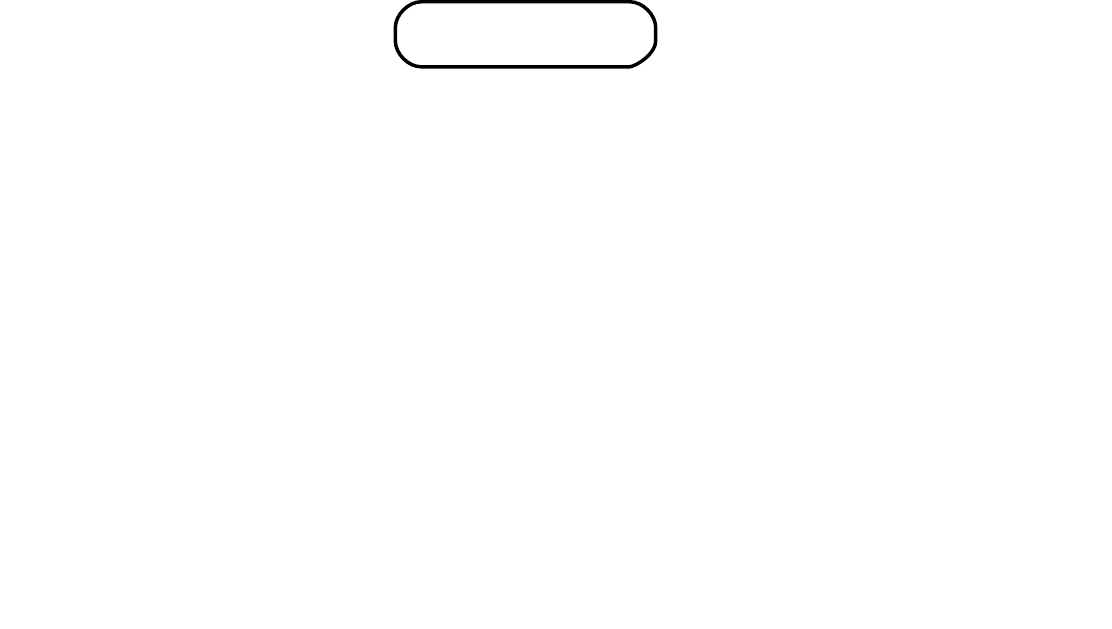
  \caption{Graphical depiction of \code{buildCommunication}.}
  \label{fig:buildCommunication}
\end{figure}

Actions are processed by two different methods, depending on their type.
In the case of communications, method \code{buildCommunication} (see
Figure~\ref{fig:buildCommunication}) computes the network resulting from executing the action, and
checks whether there exists a node in the graph containing it.
In the affirmative case, it checks whether adding an edge to that node creates a valid loop; if so,
the edge is added and the method returns \code{true}; otherwise, the method returns \code{false}.
If no such node exists, a fresh node is added with an edge to it from the current node, and
\code{buildGraph} is called recursively on the newly created node.

\begin{figure}
  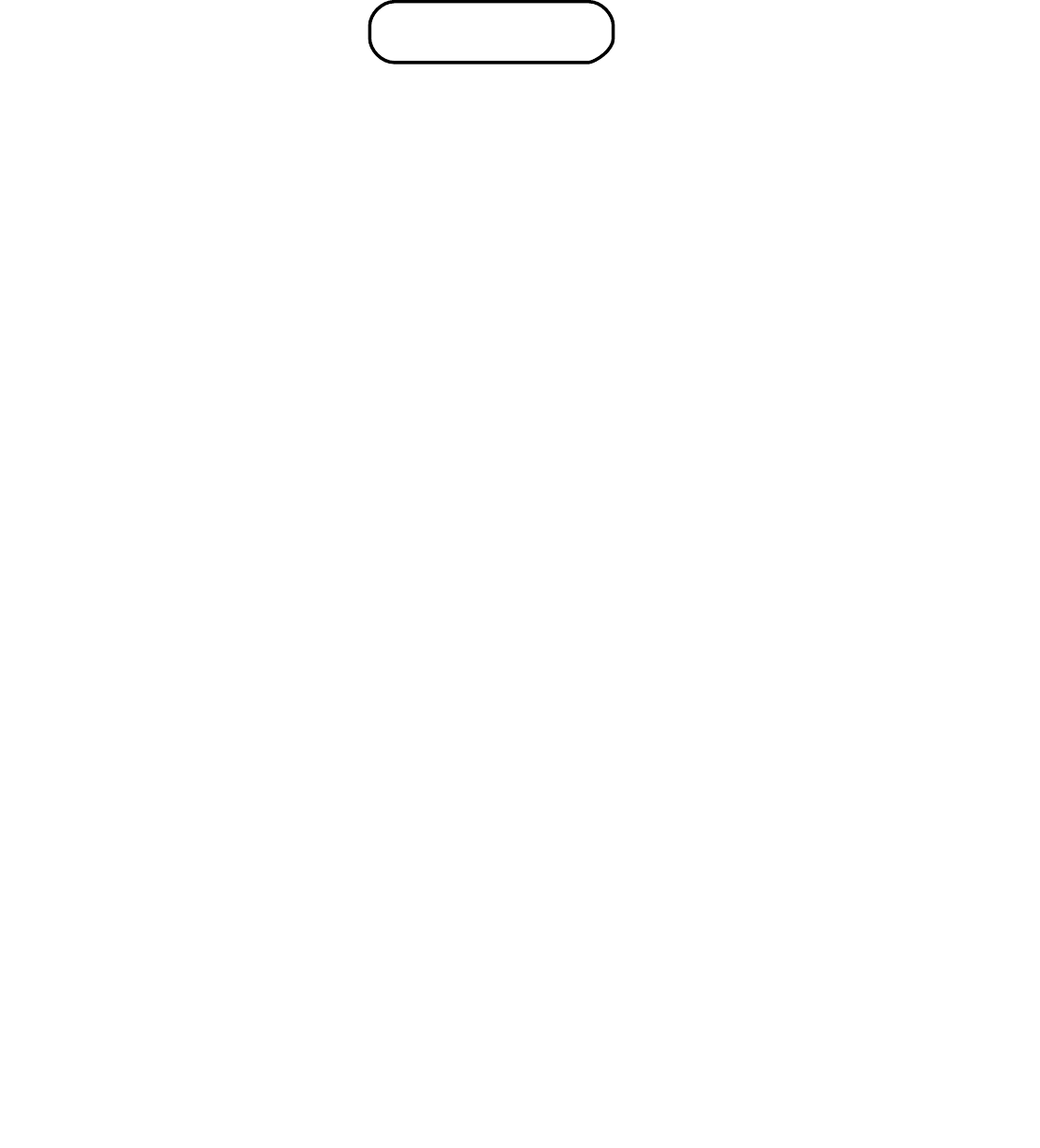
  \caption{Graphical depiction of \code{buildConditional}.}
  \label{fig:buildConditional}
\end{figure}

The case of conditionals is more involved, since two branches need to be created successfully.
Method \code{buildConditional} (Figure~\ref{fig:buildConditional}) starts by treating the
\code{then} case, much as described above, except that in case of success (by closing a loop or by
building a new node and receiving \code{true} from the recursive invocation of \code{buildGraph}) it
does not return, but moves to the \code{else} branch.
If this branch also succeeds, the method returns \code{true}; if it fails, then it returns
\code{false} and deletes all edges and nodes created in the \code{then} branch from the graph: this
step is essential for soundness of the method deciding loop validity (see
Section~\ref{sec:bad-loops}).

Edges created by \code{buildCommunication} and \code{buildConditional}
are as in Definition~\ref{defn:extr-graph}.  In the network(s) in
target node(s), we unfold exactly those procedure calls necessary for
the action labelling the edge to be executed and update the
annotations.

If the main call to \code{buildGraph} returns \code{true}, the graph
created is a valid SEG for the given network.  We then proceed to
computing a choreography according to
Definition~\ref{defn:extr-unsound}.  Method \code{unrollGraph} is
called to identify and split nodes corresponding to procedure calls.
Finally, we extract the main choreography and all procedure
definitions from the relevant nodes recursively by reading the edges
of the SEG as an abstract syntax tree, AST (method
\code{buildChoreographyBody}).

\subsection{Recognising bad loops}
\label{sec:bad-loops}

The critical part of \code{buildGraph} is deciding when a loop can be
closed.  Definition~\ref{defn:valid-seg} requires all paths that form
a loop to include a node where all processes are unmarked.  Checking
this directly is extremely inefficient, as it requires retraversing a
large part of the graph; instead, we reduce this problem to list
membership.  In order to do this, we enhance the graph structure in
different ways, so they are not simply networks anymore.  We describe
each addition below.

\paragraph{Choice-free networks.}
\label{paragraph-choice-free}
In order to best structure our explanation of our method, we first consider
the simplified case where processes do not use the conditional operator.
We construct the SEG iteratively by maintaining a
set of unexplored nodes. Whenever an unexplored node is examined,
the possible reductions lead to new terms, and, by keeping all
created nodes in a search structure, we can determine with a simple
lookup if we have created a network that already exists in the graph
we have built so far, and get a reference to that node in the SEG.
Thus, we do not recreate the node and we form a loop.

Forming a loop, we need to check if the loop contains an all-white node,
and we handle this as follows.
Since we stop our search and start backtracking when we discover a loop,
we conceptually have a path from the start node to our
current node at all times, and the path behaves in a stack-like manner.
We introduce an explicit stack as an auxiliary data structure.
Each node on the current path has a pointer to its entry on the stack.
An item on the stack contains a counter of how many white nodes can be found
further down on the stack. This information can easily be maintained
as we push and pop elements in connection with 
running the backtracking algorithm.
When we encounter a loop, we follow the pointer to the node's associated
stack item and check the counter,~$c$. The loop just found has at least one
white node if and only if the counter of the top item on the stack
is strictly greater than~$c$.

\paragraph{Choice paths.}
\label{paragraph-choice-paths}
The soundness of the strategy described above relies on the fact that,
while building the graph, no new edges are added between existing
nodes that make it possible to close a loop bypassing the edge where
the marking was erased.  This is automatically guaranteed when a
communication action is selected (the corresponding node only has one
outgoing reduction), but not in the case of conditionals.

Using the method outlined in Section~\ref{paragraph-choice-free},
following a path from an \code{else} branch, we may arrive at a node
somewhere on the \code{then} branch. Basing our decision of loop
validity on the counter may give an incorrect result, since we may
enter into some location on the \code{then} branch after the all-white node.
Thus, the counters would indicate that the loop was valid, but, in
fact, no all-white node was encountered on the \code{else} branch.

To avoid this problem, we restrict edge creation so that we can only
add an edge to an existing node if that node is a ``predecessor'' of
the current node with respect to conditionals, i.e., it was not
generated while expanding a different branch of a conditional
statement.  We do this by annotating each node with a \emph{choice
  path}: a string that represents the sequence of conditional branches
on which the node depends.  The initial node has an empty choice path,
nodes generated from a communication action inherit their parent's
choice path, and nodes generated from a conditional get their parent's
choice path appended with a $0$ or $1$ for \code{then} and \code{else}
branch, respectively.

We use choice paths in our algorithm for building a SEG
(\code{buildGraph}) as follows.  Whenever \code{buildGraph} checks
whether a node with the target network already exists in the graph, we
now additionally require the node with the target network to have a
choice path that is a prefix of the current node's (the node on which
\code{buildGraph} has been invoked); otherwise, we proceed as if no
such node exists (and create a new node).

\subsection{Well-formedness}
\label{sec:impl-wellformed}

Until this point, we have assumed networks to be well-formed (Section~\ref{sec:networks}).
While most networks that are not well-formed are not extractable (some processes are deadlocked and
therefore there is no valid SEG), this may still take a long time to detect.
Therefore, we have added an initial check that the network we are trying to extract is well-formed
(before calling \code{buildGraph}), and immediately fail in case it is not.
Having this check also allows us to assume that the network is well-formed throughout the remainder
of execution, which is relevant for some later optimisations.

\subsection{Guardedness of procedure calls}

Many previous works on process calculi require procedure calls to be \emph{guarded} (preceded by a
communication action or a conditional), in order to avoid situations such as $\procterm{X=X}{X}$.
Our language has no such restriction; however, by definition, a network containing a process whose
behaviour unfolds infinitely to a procedure call has no valid SEG: such a process would be either
livelocked in a loop or non-terminated in a leaf.

To detect these situations, our implementation includes a preprocessing check to ensure that no definition of a
procedure accessible from the main behaviour of a process can unfold to a self-call.
For example, we \emph{do} allow $\procterm{X=X}{\nil}$ and
$\procterm{\{X=Y,Y=\bsend{\pid q};X\}}{X}$, but not $\procterm{\{X=Y,Y=X\}}{X}$.
Having this previous check again simplifies the building of the graph, since we know in advance that
we cannot get into infinite loops by repeatedly unfolding a behaviour until we meet an action.

\subsection{Complexity}
Before discussing optimizations and performance on test cases, we end
this section with a discussion of the worst-case computational
complexity of our method. The starting point is the size of the AES
for an annotated network.

\begin{lemma}
\label{lem:size-aes}
The AES for an annotated network of size $n$ has at most
$e^{\frac{2n}{e}}$ vertices.
\end{lemma}
\begin{proof}
  Let $N$ be a network with $p$ processes of sizes $n_1$ through
  $n_p$, where the size of a process is the number of nodes in an
  abstract syntax tree representing the syntactical term.  Let
  $n=\sum_{i=1}^p n_i$ denote the size of $N$.

  Since recursive definitions are unfolded only when they occur at the
  top of a behaviour, a process of size $n_i$ can give rise to at most
  $n_i$ different terms when all possible reductions are considered.
  Thus, $N$ can reduce to at most $\prod_{i=1}^p n_i$ different terms.
  Since the reductions give rise to the edges in the graph, this is
  also an upper bound on the number of edges, so the graph is sparse.
  By the AM-GM inequality, $\prod_{i=1}^p n_i$ is maximised when all
  the $n_i$ are equal, where it evaluates to
  $\left(\frac{n}{p}\right)^p$.

  We now consider annotations.  Since each process is either marked or
  unmarked, there are at most $2^p$ annotations for each network,
  giving a total upper bound of $2^p(\frac{n}{p})^p=(\frac{2n}{p})^p$
  different nodes in the AES.  This expression attains its maximum
  when $p=\frac{2n}{e}$, giving the upper bound of $e^{\frac{2n}{e}}$
  nodes in the AES.
\end{proof}

Next we consider the extraction of the enhanced SEG from the AES.
\begin{theorem}
  \label{thm:extract-complex}
  Extraction from a network of size $n$ with $c$ conditionals terminates in time $O(2^c n e^{\frac{2n}{e}})$.
\end{theorem}
\begin{proof}
  For networks without conditionals, we develop a network of size at
  most $e^{\frac{2n}{e}}$, as outlined in
  Sections~\ref{paragraph-choice-free} and bounded in
  Lemma~\ref{lem:size-aes}. However, adding the choice path as part of
  the node identity, as outlined in
  Section~\ref{paragraph-choice-paths}, the number of possible
  different nodes is increased by a factor $2^c$, representing all
  possible choice paths, where $c$ is the number of conditionals in
  the network -- which is of course (much) smaller than~$n$. Look-up for
  a node to check if it is new has time complexity worst-case
  logarithmic in the size of the set of nodes, using any standard dictionary
  implementation, i.e., $O(\log(2^ce^{\frac{2n}{e}}))\subseteq O(n)$,
  plus a check for term identity which is also $O(n)$.  Clearly
  maintaining the auxiliary stack takes constant time in each step,
  and with the stack available, we can check for bad loops in constant
  time as well. Thus, the overall time complexity is
  $O(n 2^c e^{\frac{2n}{e}})$.
\end{proof}

As mentioned in the introduction, our method avoids the
factorial time complexity of previous work.  Exponential time is
better than factorial, but we may perform even better in
practice.  Algorithmically, all the required work stems from traversals
of the AES, so any reduction in its (explored) size will lead to
proportional runtime improvements.  We point out that in the
algorithms proposed above, instead of first computing
the entire AES and then a valid SEG, we compute the relevant parts
of the AES lazily as we need them. Thus, parts of the AES that are never
explored while computing a valid SEG are never generated.

\section{Extensions and optimisations}
\label{sec:optim}

We now discuss some extensions and optimisations to the original algorithm.
Some of these changes aim at making make the implementation more efficient in situations that may
occur often enough to warrant consideration; others extend the domain of extractable choreographies,
and were motivated by practical applications.

\subsection{Parallelisation}
\label{sec:impl-parallel}

In our first testing phase (see Section~\ref{sec:eval}), we took the benchmarks from~\cite{LTY15}
and wrote them as networks.
This translation was done by hand, ensuring that the network represented the same protocol as the
communicating automata in the original work.
Of these, 3 benchmarks (\emph{alternating 2-bit}, \emph{alternating 3-bit} and \emph{TPMContract})
were not implemented.
The first uses group communications, an extension described in~\cite{CLM17} that is not implemented;
the two last require local threads, and are not representable in our formalism.
(We discuss this in the conclusions.)

Several benchmarks are parallel compositions of two instances of the same network.
They exhibit a very high degree of parallelism, visibly slowing down extraction.
However, very simple static analysis can easily improve performance in such instances.
We define the network's communication graph as the undirected graph whose nodes are processes, and
where there is an edge between $\pid p$ and $\pid q$ if they ever interact.
The connected components of this graph can be extracted independently, and the choreographies
obtained composed in parallel at the end.

Theoretically, this requires adding a parallel composition constructor at the top level of a
choreography, which is straightforward.
In practice, this trivial preprocessing drastically reduces the computation time: for the (very
small) benchmarks from~\cite{LTY15}, doubling the size of the network already corresponds to a
increase in computation time of up to $35$ times, while splitting the network in two and extracting
each component in sequence keeps this factor under $2$, since the independent components can be
extracted in parallel.

We report our empirical evaluation in Table~\ref{tab:nobuko}.
These numbers are purely indicative: due to the very small size of these examples, we did not
attempt to make a very precise evaluation.
We measured the extraction time for all benchmarks, and computed the ratio between each benchmark
containing a duplicate network and the non-duplicated one.
This was done with the original, sequential, algorithm, and with the parallelised one.
All execution times were averaged over three runs.
The values themselves are not directly comparable to those from~\cite{LTY15}, since the network
implementations are substantially different, but the ratios show the advantages of our approach:
even without parallelisation, our ratios are substantially lower, in line with the better
asymptotical complexity of our method shown in~\cite{CLM17}.
(Note that the examples from~\cite{LTY15} where the ratio is lowest are the smallest ones, where the
execution time is dominated by the setup and command-line invocation of the different programs
used.)

\begin{table}
  \centering
  \caption{Empirical evaluation of the effect of parallelising extraction (all times in ms).}
  \label{tab:nobuko}

  \begin{tabular}{l@{\hspace{1em}}rrr@{\hspace{1em}}rrr@{\hspace{1em}}rrr}
    \toprule
    Test name & \multicolumn3c{sequential} & \multicolumn3c{parallel} & \multicolumn3c{from~\cite{LTY15}} \\
    & single & double & ratio & single & double & ratio & single & double & ratio \\ \midrule
    Bargain &
    1.7 & 10.0 & \bf5.88 &
    3.0 & 3.7 & \bf1.23 &
    103 & 161 & \bf1.56 \\
    Cloud system &
    8.3 & 83.0 & \bf10.0 &
    8.6 & 8.3 & \bf0.96 &
    140 & 432 & \bf3.08 \\
    Filter collaboration &
    4.0 & 123.3 & \bf30.83 &
    5.0 & 4.7 & \bf0.93 &
    118 & 178 & \bf1.51 \\
    Health system &
    6.0 & 80.3 & \bf13.39 &
    7.3 & 11.7 & \bf1.59 &
    17 & 1702 & \bf100.12 \\
    Logistic &
    1.0 & 34.7 & \bf34.70 &
    5.3 & 16.7 & \bf3.14 &
    276 & 2155 & \bf7.81 \\
    Running example &
    7.7 & 143.3 & \bf18.61 &
    5.7 & 6.7 & \bf1.17 &
    184 & 22307 & \bf121.23 \\
    Sanitary agency &
    6.0 & 61.0 & \bf10.17 &
    8.0 & 7.3 & \bf0.92 &
    241 & 3165 & \bf13.13 \\
    \bottomrule
  \end{tabular}
\end{table}

\subsection{Extraction strategies}
\label{sec:impl-strategy}

The performance of our implementation depends on the choice of the network action, in cases where
there are several possible options: expanding a communication generates one descendant node, but
expanding a conditional generates two descendant nodes that each need to be processed.
On the other hand, if the choreography contains cyclic behaviour, different choices of actions may
impact the size of the extracted loops (and thus also execution time).

In order to control these choices, we define \emph{execution strategies}: heuristics that guide the
choice of the next action to pick.
Strategies either take into account the syntactic type of the action (e.g., prioritise interactions)
or the semantics of bad loops (prioritise unmarked processes), or combine them with different
priorities (prioritise unmarked processes and break ties by preferring interactions).
We also include a basic strategy that picks a random action.

All strategies are implemented in the same way: in \code{buildGraph}, we choose from the list of possible actions
that the network in the current node according to the chosen criterion.

Our implementation includes the following strategies.
The abbreviations in parenthesis are used in captions of graphics.
\begin{description}
\item[\texttt{Random (R)}] Choose a random action.
\item[\texttt{LongestFirst (L)}] Prioritise the process with the largest body.
\item[\texttt{ShortestFirst (S)}] Prioritise the process with the smallest body.
\item[\texttt{InteractionsFirst (I)}] Prioritise interactions.
\item[\texttt{ConditionalsFirst (C)}] Prioritise conditionals.
\item[\texttt{UnmarkedFirst (U)}] Prioritise actions involving unmarked processes.
\item[\texttt{UnmarkedThenInteractions (UI)}] Prioritise actions involving unmarked processes, and
  as secondary criterion prioritise interactions.
\item[\texttt{UnmarkedThenSelections (US)}] Prioritise unmarked processes, as a secondary criterion
  prioritise selections, and afterwards value communications.
\item[\texttt{UnmarkedThenConditionals (UC)}] Prioritise unmarked processes, and as secondary
  criterion prioritise conditionals.
\item[\texttt{UnmarkedThenRandom (UR)}] Prioritise unmarked processes, in random order.
\end{description}

We remark that \texttt{UnmarkedFirst} and \texttt{UnmarkedThenRandom} are different strategies:
\texttt{UnmarkedFirst} does not distinguish among actions involving unmarked processes, so they come
in the order of the processes involved in the network.
By contrast, \texttt{UnmarkedThenRandom} chooses randomly from the list of possible actions, in
principle contributing towards more fairness among processes.

From the results in the next section, we see that \texttt{LongestFirst} and \texttt{ShortestFirst}
perform significantly worse than all other strategies, while \texttt{Random} and
\texttt{UnmarkedFirst} in general give the best results.
However, we remark that comparing the performances of different strategies was not an objective of
this work, as it would require a dedicated test suite.
We leave it as interesting future work.

\subsection{Livelocks}
\label{sec:impl-services}

Several examples in~\cite{LTY15} include processes that offer a service, and as such may be inactive throughout a part (or the whole) of execution.
This is the case in our Example~\ref{ex:aes}: process $\pid r$ provides a value to $\pid q$ whenever it is needed, but $\pid q$ might stop requesting values during execution.
Our extraction algorithm does not allow for this behaviour: when a loop is closed, every process must either be terminated or reduce inside the loop.

In order to allow for services, we added a parameter to the extraction method containing a list of
services (in our example, $\pid r$), which are not required to reduce inside loops.
Intuitively, we ignore the annotations in these processes when deciding whether a loop is valid.
In the implementation, these processes are marked initially, and are not unmarked when the marking
is erased.

\subsection{Clever backtracking}

Our strategy of building the SEG in a depth-first fashion requires that, on failure, we backtrack
and explore different possible actions.
This leads to a worst-case behaviour where all possible execution paths need to be explored, in the
case that no choreography can be extracted from the original network.
However, a closer look at why a particular branch leads to deadlock allows us to avoid backtracking
in some instances: network execution is confluent, so if we reach a deadlocked state, then every
possible execution reaches such a state, and extraction must fail.
It is only when extraction fails because of attempting to close an invalid loop that backtracking is
required.

To implement this refinement, the return type of all methods that try to build an edge of the SEG
was changed to a $3$-element set.
If a method succeeds, it returns \m{ok} (corresponding to \code{true}); if it fails due to reaching
a deadlock, it returns \m{fail} (corresponding to \code{false}); and if it fails due to trying to
close an invalid loop, it returns \m{badloop}.
In recursive calls, these values are treated as follows:
\begin{itemize}
\item if the caller is processing a communication or the \code{else} branch of a conditional, they
  are propagated upwards;
\item if the caller is processing the \code{then} branch of a conditional, \m{fail} and \m{badloop}
  are propagated upwards, while \m{ok} signals that the \code{else} branch can now be treated.
\end{itemize}

For \code{buildGraph}, a method call returning \m{ok} or \m{fail} is also propagated upwards, while
\m{badloop} signals that a different possible action should be tried.
If all possible actions return \m{badloop}, then \code{buildGraph} returns \m{fail}.
This is sound: due to confluence, any action that could have been executed before that would make it
possible to close a loop from this node can also be executed from this node.

This optimisation is crucial to get a practical implementation in the case of \emph{unextractable}
networks.
Most of the failure tests (Section~\ref{sec:testing-bad}) did not terminate before this change,
while they now fail in time comparable to that of success.

\section{Practical evaluation}
\label{sec:eval}

In order to evaluate the performance of our implementation, we developed a three-stage plan.

\begin{enumerate}[{Phase }1.]
\item We focused on the test cases from~\cite{LTY15}, in order to ensure that our tool covered at
  least those cases.
  Since these cases are simple, we verified their correctness by hand.
\item We generated 1050 random choreographies and their projections by varying four different
  parameters (see details below), and applied our tool to the projected networks.
  In this way, we tested whether we can extract networks that are direct projections of
  choreographies -- these should correspond to the majority of (extractable) practical applications.
  Soundness can be checked by testing that the extracted choreography is bisimilar to the original
  one.
\item We proposed a model for the typical changes (correct or incorrect) introduced when a
  programmer modifies a process directly, and tried to extract choreographies from the resulting
  networks.
  This yielded information about how quickly our program fails when a network is unextractable; as a
  side result, we also got information about how often some types of protocol errors can slip through
  undetected, that is, the network is still extractable, but it implements a different protocol than the
  original.
\end{enumerate}

We deliberately did \emph{not} generate any networks directly.
We claim that such tests are not very meaningful for two reasons: first, they do not correspond to 
realistic scenarios; second, randomly generated networks are nearly always unextractable.
We believe our test suite is comprehensive enough to model most situations with practical relevance.

All tests reported in this section were performed on a computer running Arch Linux, kernel version 5.14.8, with an AMD Ryzen 9 3950x as CPU and 50 GB RAM as available memory for the Java Virtual Machine.

\subsection{Comparison with the literature}
Our first testing phase used the benchmarks from~\cite{LTY15}.
As described in Section~\ref{sec:impl-parallel}, the networks corresponding to those examples were
written by hand.
These tests were done simply as a proof-of-concept, as their simplicity means that the measured
execution times are extremely imprecise.
As discussed earlier, three test cases were not implementable; all others succeeded.
The results (using strategy \texttt{InteractionsFirst}) are reported in Table~\ref{tab:nobuko}.

\subsection{Reverse projection}
In the second phase, we generate large-scale tests to check the scalability of our implementation.
Our tests consist of randomly-generated choreographies characterised by four parameters: number of
processes, total number of actions, number of those actions that should be conditionals, and a number
of procedures.

Then, we generate ten choreographies for each set of parameters as follows: first, we determine how
many actions and conditionals each procedure definition (including \code{main}) should have by
uniformly partitioning the total number of actions and conditionals.
Then we generate the choreography sequentially by randomly choosing the type of the next action so
that the probability distribution of conditional actions within each procedure body is uniform.
For each action, we randomly choose the process(es) involved, again with uniform distribution, and
assigned fresh values to any expression or label involved.
At the end, we randomly choose whether to end with termination or a procedure call.
Finally, we apply rules for swapping conditionals (rules \rname{C}{Cond-Eta} and \rname{C}{Cond-Cond} from Figure~\ref{fig:cc_precongr}) to obtain inefficient representations of choreographies where code is duplicated in both branches of a conditional. (This actually increases the number of conditionals in a choreography from at most $50$ to over $8000$ in some cases.)

This method may generate choreographies with dead code (if some procedures are never called).
Therefore there is a post-check that determines whether every procedure is reachable from
\code{main} (possibly dependent on the results of some conditional actions); if this is not the
case, the choreography is rejected, and a new one is generated.

A randomly generated choreography with conditional actions is typically unprojectable, so we amend
it (see~\cite{CM20}) to make it projectable.
In general, this increases the size of the choreography.
Finally, we apply projection to obtain the networks for our second test suite.

\begin{table}
  \caption{Parameters for the choreographies generated for testing.}
  \label{tab:chor-params}
  \centering

  \begin{tabular}{ccrrrrr}
    \toprule 
    Test set & parameter & size & processes & \code{if}s & \code{def}s & \#\ tests \\ \midrule
    size & $k\in[1..42]$ & $50k$ & $6$ & $0$ & $0$ & $420$ \\
    processes & $k\in[1..20]$ & $500$ & 5$k$ & $0$ & $0$ & $200$ \\
    ifs (finite) & $k\in[1..4]$ & $50$ & $6$ & $10k$ & 0 & $40$ \\
    ifs (varying procedures) & $\tuple{j,k}\in[0..5]\times[0..3]$ & $200$ & $5$ & $j$ & 5$k$ & $240$ \\
    procedures (fixed ifs) & $k\in[1..15]$ & $20$ & $5$ & $8$ & $k$ & $150$ \\ \midrule
    total &&&&&& $1050$ \\ \bottomrule
  \end{tabular}
\end{table}

The parameters for generation are given in Table~\ref{tab:chor-params}.
The upper bounds were determined by our hardware limitations.
Four of the generated files contained tests that were too large to extract, and were removed from
the final test set.

\paragraph{Results.}
We report on the most interesting tests.
The first test shows that, predictably, for choreographies consisting of only communications, the
extraction time is nearly directly proportional to the network size (with a small overhead
  from needing to work with larger objects), except when using strategies that need to compute the
size of each process term.
We could enrich the networks with this information in order to make these strategies more efficient,
but since they perform poorly in general, we did not pursue this approach.

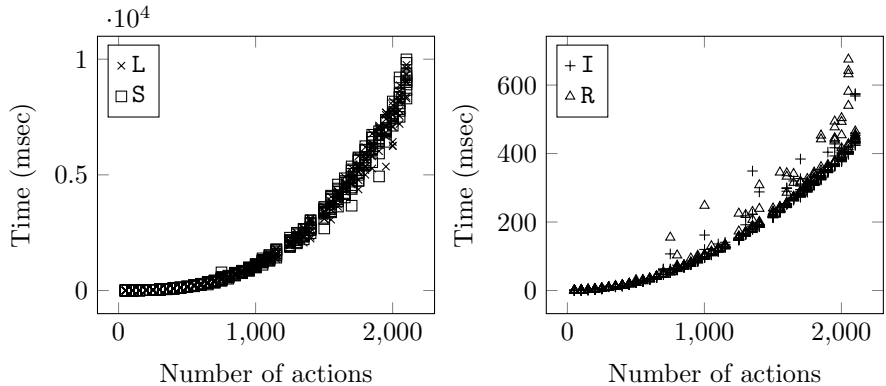
\begin{figure}
  \begin{tikzpicture}
    \begin{axis}[
        width=.5\textwidth,
        height=15em,
        xlabel=Number of actions,
        ylabel=Time (msec),
        legend pos=north west
      ]
      \addplot[only marks,mark=x]
      table[x=numberOfActions,y=time(msec)-LongestFirst,col sep=tab]{stats-comms-only.csv};
      \addlegendentry{\texttt{L}}
      \addplot[only marks,mark=square]
      table[x=numberOfActions,y=time(msec)-ShortestFirst,col sep=tab]{stats-comms-only.csv};j
      \addlegendentry{\texttt{S}}
    \end{axis}
  \end{tikzpicture}
  \begin{tikzpicture}
    \begin{axis}[
        width=.5\textwidth,
        height=15em,
        xlabel=Number of actions,
        ylabel=Time (msec),
        legend pos=north west
      ]
      \addplot[only marks,mark=+]
      table[x=numberOfActions,y=time(msec)-InteractionsFirst,col sep=tab]{stats-comms-only.csv};
      \addlegendentry{\texttt{I}}
      \addplot[only marks,mark=triangle]
      table[x=numberOfActions,y=time(msec)-Random,col sep=tab]{stats-comms-only.csv};
      \addlegendentry{\texttt{R}}
    \end{axis}
  \end{tikzpicture}
  
  \caption{Execution time vs.\ length for networks consisting only of communications.
    The omitted strategies essentially perform as \texttt{InteractionsFirst}, since there are no
    other types of actions and no recursive procedures.}
  \label{fig:results-test-1}
\end{figure}

The second test is similar, but varying the number of processes (which makes for a greater number of
possible actions at each step) while keeping the size constant.
Our results show that execution time grows linearly with the number of processes for
\texttt{InteractionFirst} and \texttt{Random}.
The behaviour of \texttt{LongestFirst} and \texttt{ShortestFirst} is more interesting, as the time for computing the length of the behaviours dominates for small numbers of processes.

\begin{figure}
  \begin{tikzpicture}
    \begin{axis}[
        width=.5\textwidth,
        height=15em,
        xlabel=Number of processes,
        ylabel=Time (msec),
        legend pos=north west
      ]
      \addplot[only marks,mark=x]
      table[x=numberOfProcesses,y=time(msec)-LongestFirst,col sep=tab]{stats-increasing-processes.csv};
      \addlegendentry{\texttt{L}}
      \addplot[only marks,mark=square]
      table[x=numberOfProcesses,y=time(msec)-ShortestFirst,col sep=tab]{stats-increasing-processes.csv};
      \addlegendentry{\texttt{S}}
    \end{axis}
  \end{tikzpicture}
  \begin{tikzpicture}
    \begin{axis}[
        width=.5\textwidth,
        height=15em,
        xlabel=Number of processes,
        ylabel=Time (msec),
        legend pos=north west
      ]
      \addplot[only marks,mark=+]
      table[x=numberOfProcesses,y=time(msec)-InteractionsFirst,col sep=tab]{stats-increasing-processes.csv};
      \addlegendentry{\texttt{I}}
      \addplot[only marks,mark=triangle]
      table[x=numberOfProcesses,y=time(msec)-Random,col sep=tab]{stats-increasing-processes.csv};
      \addlegendentry{\texttt{R}}
    \end{axis}
  \end{tikzpicture}
  
  \caption{Execution time vs.\ number of processes, with constant total number of actions.}
  \label{fig:results-test-2}
\end{figure}
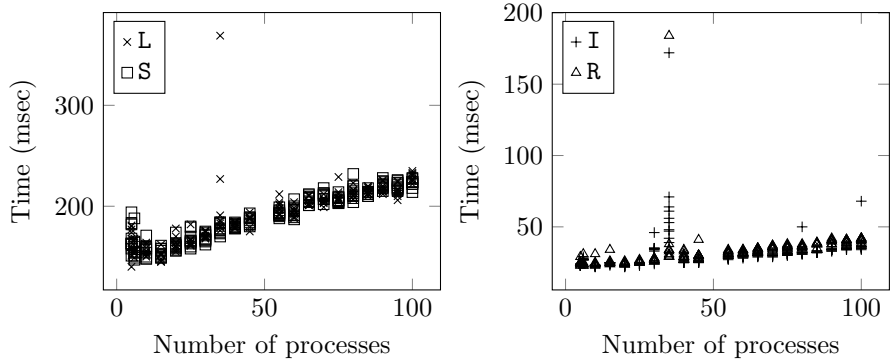

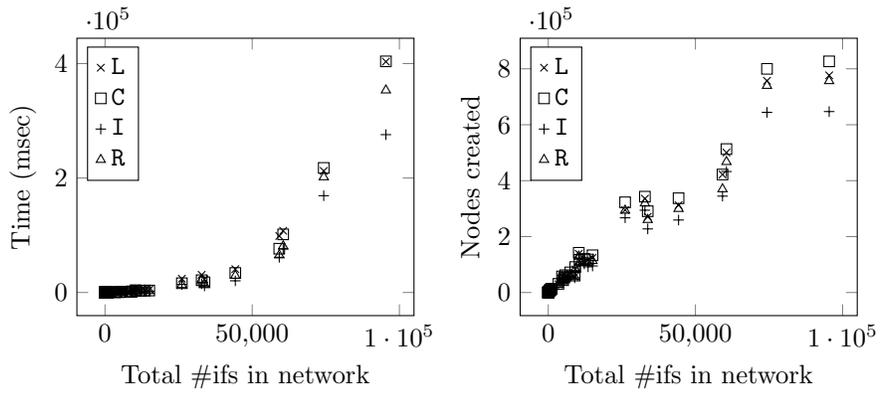
\begin{figure}
  \begin{tikzpicture}
    \begin{axis}[
        width=.5\textwidth,
        height=15em,
        xlabel=Total \#ifs in network,
        ylabel=Time (msec),
        legend pos=north west,
        scaled x ticks=false
      ]
      \addplot[only marks,mark=x]
      table[x=numberOfConditionals,y=time(msec)-LongestFirst,col sep=tab]{stats-increasing-ifs-no-recursion.csv};
      \addlegendentry{\texttt{L}}
      \addplot[only marks,mark=square]
      table[x=numberOfConditionals,y=time(msec)-ConditionsFirst,col sep=tab]{stats-increasing-ifs-no-recursion.csv};
      \addlegendentry{\texttt{C}}
      \addplot[only marks,mark=+]
      table[x=numberOfConditionals,y=time(msec)-InteractionsFirst,col sep=tab]{stats-increasing-ifs-no-recursion.csv};
      \addlegendentry{\texttt{I}}
      \addplot[only marks,mark=triangle]
      table[x=numberOfConditionals,y=time(msec)-Random,col sep=tab]{stats-increasing-ifs-no-recursion.csv};
      \addlegendentry{\texttt{R}}
    \end{axis}
  \end{tikzpicture}
  \begin{tikzpicture}
    \begin{axis}[
        width=.5\textwidth,
        height=15em,
        xlabel=Total \#ifs in network,
        ylabel=Nodes created,
        legend pos=north west,
        scaled x ticks=false
      ]
      \addplot[only marks,mark=x]
      table[x=numberOfConditionals,y=nodes-LongestFirst,col sep=tab]{stats-increasing-ifs-no-recursion.csv};
      \addlegendentry{\texttt{L}}
      \addplot[only marks,mark=square]
      table[x=numberOfConditionals,y=nodes-ConditionsFirst,col sep=tab]{stats-increasing-ifs-no-recursion.csv};
      \addlegendentry{\texttt{C}}
      \addplot[only marks,mark=+]
      table[x=numberOfConditionals,y=nodes-InteractionsFirst,col sep=tab]{stats-increasing-ifs-no-recursion.csv};
      \addlegendentry{\texttt{I}}
      \addplot[only marks,mark=triangle]
      table[x=numberOfConditionals,y=nodes-Random,col sep=tab]{stats-increasing-ifs-no-recursion.csv};
      \addlegendentry{\texttt{R}}
    \end{axis}
  \end{tikzpicture}
  
  \caption{Execution time (left) and number of nodes (right) vs.\ total number of conditionals, for
    networks consisting only of conditionals.
    The omitted strategies essentially perform as \texttt{InteractionsFirst} or
    \texttt{ConditionalsFirst}.}
  \label{fig:results-test-3}
\end{figure}

The third test introduces conditionals.
Our results show that execution time varies with the \emph{total} number of conditionals in the
network, rather than with the number of conditionals in each process.
Figure~\ref{fig:results-test-3} (left) exhibits the worst-case exponential behaviour of our
algorithm, and also suggests that delaying conditionals is in general a better strategy.
Figure~\ref{fig:results-test-3} (right) shows the number of nodes created in the SEG, illustrating
that execution time is not directly proportional to this value.

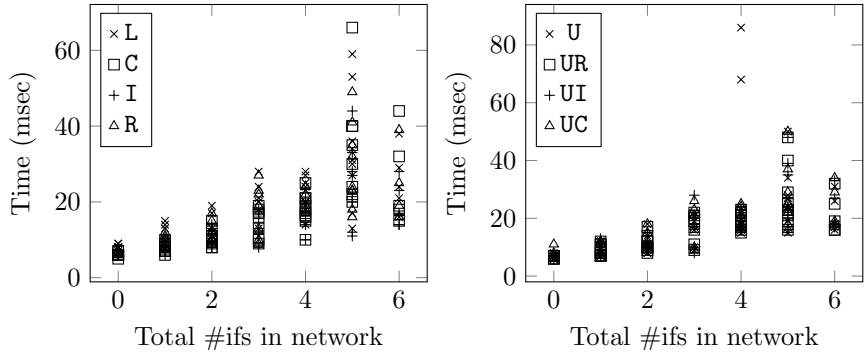
\begin{figure}
  \begin{tikzpicture}
    \begin{axis}[
        width=.5\textwidth,
        height=15em,
        xlabel=Total \#ifs in network,
        ylabel=Time (msec),
        legend pos=north west,
        scaled x ticks=false
      ]
      \addplot[only marks,mark=x]
      table[x=numberOfConditionals,y=time(msec)-LongestFirst,col sep=tab]{stats-increasing-ifs-with-recursion.csv};
      \addlegendentry{\texttt{L}}
      \addplot[only marks,mark=square]
      table[x=numberOfConditionals,y=time(msec)-ConditionsFirst,col sep=tab]{stats-increasing-ifs-with-recursion.csv};
      \addlegendentry{\texttt{C}}
      \addplot[only marks,mark=+]
      table[x=numberOfConditionals,y=time(msec)-InteractionsFirst,col sep=tab]{stats-increasing-ifs-with-recursion.csv};
      \addlegendentry{\texttt{I}}
      \addplot[only marks,mark=triangle]
      table[x=numberOfConditionals,y=time(msec)-Random,col sep=tab]{stats-increasing-ifs-with-recursion.csv};
      \addlegendentry{\texttt{R}}
    \end{axis}
  \end{tikzpicture}
  \begin{tikzpicture}
    \begin{axis}[
        width=.5\textwidth,
        height=15em,
        xlabel=Total \#ifs in network,
        ylabel=Time (msec),
        legend pos=north west,
        scaled x ticks=false
      ]
      \addplot[only marks,mark=x]
      table[x=numberOfConditionals,y=time(msec)-UnmarkedFirst,col sep=tab]{stats-increasing-ifs-with-recursion.csv};
      \addlegendentry{\texttt{U}}
      \addplot[only marks,mark=square]
      table[x=numberOfConditionals,y=time(msec)-UnmarkedThenRandom,col sep=tab]{stats-increasing-ifs-with-recursion.csv};
      \addlegendentry{\texttt{UR}}
      \addplot[only marks,mark=+]
      table[x=numberOfConditionals,y=time(msec)-UnmarkedThenInteractions,col sep=tab]{stats-increasing-ifs-with-recursion.csv};
      \addlegendentry{\texttt{UI}}
      \addplot[only marks,mark=triangle]
      table[x=numberOfConditionals,y=time(msec)-UnmarkedThenConditions,col sep=tab]{stats-increasing-ifs-with-recursion.csv};
      \addlegendentry{\texttt{UC}}
    \end{axis}
  \end{tikzpicture}
  
  \caption{Execution time vs.\ total number of conditionals for networks including 5 recursive procedures.}
  \label{fig:results-test-4}
\end{figure}

The behaviour when recursive procedures also occur is shown in Figure~\ref{fig:results-test-4},
where we fix the number of procedures to 5.

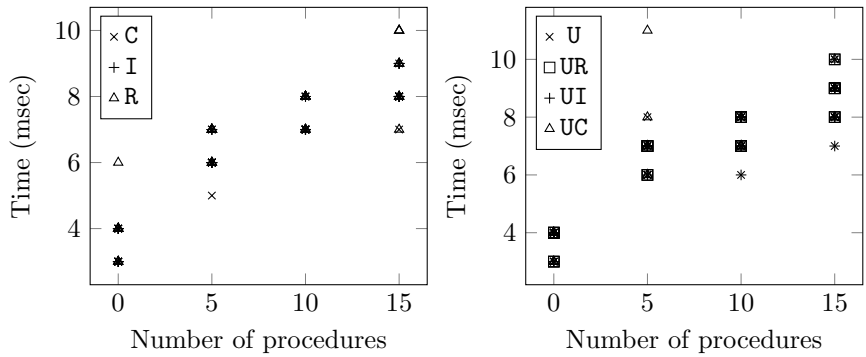
\begin{figure}
  \begin{tikzpicture}
    \begin{axis}[
        width=.5\textwidth,
        height=15em,
        xlabel=Number of procedures,
        ylabel=Time (msec),
        legend pos=north west
      ]
      \addplot[only marks,mark=x]
      table[x=numberOfProcedures,y=time(msec)-ConditionsFirst,col sep=tab]{stats-increasing-procedures-no-ifs.csv};
      \addlegendentry{\texttt{C}}
      \addplot[only marks,mark=+]
      table[x=numberOfProcedures,y=time(msec)-InteractionsFirst,col sep=tab]{stats-increasing-procedures-no-ifs.csv};
      \addlegendentry{\texttt{I}}
      \addplot[only marks,mark=triangle]
      table[x=numberOfProcedures,y=time(msec)-Random,col sep=tab]{stats-increasing-procedures-no-ifs.csv};
      \addlegendentry{\texttt{R}}
    \end{axis}
  \end{tikzpicture}
  \begin{tikzpicture}
    \begin{axis}[
        width=.5\textwidth,
        height=15em,
        xlabel=Number of procedures,
        ylabel=Time (msec),
        legend pos=north west
      ]
      \addplot[only marks,mark=x]
      table[x=numberOfProcedures,y=time(msec)-UnmarkedFirst,col sep=tab]{stats-increasing-procedures-no-ifs.csv};
      \addlegendentry{\texttt{U}}
      \addplot[only marks,mark=square]
      table[x=numberOfProcedures,y=time(msec)-UnmarkedThenRandom,col sep=tab]{stats-increasing-procedures-no-ifs.csv};
      \addlegendentry{\texttt{UR}}
      \addplot[only marks,mark=+]
      table[x=numberOfProcedures,y=time(msec)-UnmarkedThenInteractions,col sep=tab]{stats-increasing-procedures-no-ifs.csv};
      \addlegendentry{\texttt{UI}}
      \addplot[only marks,mark=triangle]
      table[x=numberOfProcedures,y=time(msec)-UnmarkedThenConditions,col sep=tab]{stats-increasing-procedures-no-ifs.csv};
      \addlegendentry{\texttt{UC}}
    \end{axis}
  \end{tikzpicture}
  
  \caption{Execution time vs.\ number of procedures, no conditionals.}
  \label{fig:results-test-5}
\end{figure}

The final tests introduce variations in the number of procedures.
The results of these tests are too complex to allow for immediate conclusions.
Figure~\ref{fig:results-test-5} shows what happens when we vary the number of procedures for
choreographies without conditionals.
Although the number of procedures potentially influences the number of loops in the AES, this
dependency is likely too complex to be visible in the test results.

When we vary the number of procedures in more complex scenarios, the picture is even less clear, and
we omit a discussion of these results.

\paragraph{Correctness.}
In order to obtain confirmation of the correctness of our algorithm and its implementation, we
performed an additional verification at this point.
We implemented a naive similarity checker that tests whether a choreography $C_1$ can simulate
another choreography $C_2$ as follows: we keep a set of pairs $R$, initially containing only the
pair \tuple{C_1,C_2}.
At each step, we choose a pair \tuple{C,C'} from $R$ and compute all actions $\alpha$ and
choreographies $C_\alpha$ such that $C$ can reach $C_\alpha$ by executing $\alpha$.
For each such action $\alpha$, we check that $C'$ can execute $\alpha$, compute the resulting
choreography $C'_\alpha$, and add the pair \tuple{C_\alpha,C'_\alpha} to $R$.
If $C'$ cannot execute $\alpha$, the checker returns \code{false}.
When all pairs in $R$ have been processed, the checker returns \code{true}.

We then check, for each test, that the original choreography and the one obtained by extraction
can simulate each other.

\begin{lemma}
  If $C_1$ and $C_2$ can simulate each other, then there is a bisimulation between $C_1$ and $C_2$.
\end{lemma}
\begin{proof}
  We first observe that the final set $R$ computed by the algorithm is always the same, regardless
  of the order in which pairs are picked.
  
  Let $R_{12}$ and $R_{21}$ be the sets built when checking that $C_2$ simulates $C_1$ and that
  $C_2$ simulates $C_1$, respectively.
  We show by induction on the construction of $R_{12}$ that $R_{12}^{-1}\subseteq R_{21}$.
  Initially this holds, since $R_{12}=\{\tuple{C_1,C_2}\}$ and \tuple{C_2,C_1} is initially in
  $R_{21}$.
  Suppose $\tuple{C,C'}\in R_{12}$ is selected for processing.
  By induction hypothesis, $\tuple{C',C}\in R_{21}$.
  For every $\alpha$ such that $C$ can execute $\alpha$ and move to $C_\alpha$, there is a unique
  choreography $C'_\alpha$ such that $C'$ can execute $\alpha$ and move to $C'_\alpha$.
  Therefore, in the step where \tuple{C',C} is selected from $R_{21}$, every such pair
  \tuple{C'_\alpha,C_\alpha} is added to $R_{21}$, hence it is in the final set.
  Thus, after extending $R_{12}$ with all the pairs obtained from \tuple{C,C'}, the thesis still holds.

  By reversing the roles of $C$ and $C'$, we also establish that $R_{21}^{-1}\subseteq R_{12}$.
  Therefore $R_{12}=R_{21}^{-1}$.
  It then follows straightforwardly that $R_{12}$ is a bisimulation between $C_1$ and $C_2$.
\end{proof}

Given that bisimulation is in general undecidable and that we did not make any effort to make a
clever implementation, our program often runs out of resources without terminating.
Still, it finished in about 5\%\ of the tests (those of smaller size), always with a positive
result.
While this may not sound impressive, it is unlikely that errors in the implementation would only
show up in larger tests, and this result increases our confidence in the soundness of the
implementation.

\subsection{Fuzzer and unroller}
\label{sec:testing-bad}

In the third testing phase, we changed the networks obtained by choreography projection using two
different methods.
The first method (the \emph{fuzzer}) applies transformations that are semantically incorrect, and
typically result in unextractable networks (modelling programmer errors).
The second method (the \emph{unroller}) applies transformations that are semantically correct, and
result in networks that are bisimilar to the original and should be extractable (modelling
alternative implementations of the same protocol).

\paragraph{The fuzzer.}
For the fuzzer, we considered the following transformations: adding an action; removing an action; and
switching the order of two actions.
The first two always result in an unextractable network, whereas the latter may still give an
extractable network that possibly implements a different protocol.

Our fuzzer takes two parameters $d$ and $s$, randomly chooses one process in the network, deletes
$d$ actions in its definition and switches $s$ actions with the following one.
The probability distribution of deletions and swaps is uniform (all actions have the same
probability of being deleted of swapped).
We made the following conventions: deleting a conditional preserves only the \code{then} branch;
deleting a branching term preserves only the first branch offered; swapping a conditional or
branching with the next action switches it with the first action in the \code{then}/first branch;
and swapping the last action in a behaviour with the next one amounts to deleting that action.
Deleting a conditional results in an extractable network that implements a subprotocol of the
original one, while other deletions yield unextractable networks.
Exchanges of communication actions may yield extractable networks, but with a different extracted
choreography; all other types of exchanges break extractability.

We did not implement adding a random action, as this is covered in our tests: adding an unmatched
send from \m{p} to \m{q} can be seen as removing a receive at \m{q} from \m{p} from a choreography
that includes that additional communication.
We restricted fuzzing to one process only since in practice we can assume that processes are changed
one at a time.
We applied three different versions of fuzzing to all our networks: one swap; one deletion; and two
swaps and two deletions.
The results are summarised in Table~\ref{tab:fuzzing}.

\begin{table}
  \caption{Extracting fuzzed networks: for each strategy we report on the percentage of
    unextractable networks (\%) and the average and median times to fail in those cases (ms).
  We highlight the best and worst values in each column in green and red, respectively.}
  \label{tab:fuzzing}
  \centering
  \begin{tabular}{c|rrr|rrr|rrr}
    \toprule
    Strategy
    & \multicolumn3{c|}{$d=0$, $s=1$}
    & \multicolumn3{c|}{$d=1$, $s=0$}
    & \multicolumn3{c}{$d=2$, $s=2$}
    \\
    & \% & avg & med
    & \% & avg & med
    & \% & avg & med
    \\ \midrule
    \texttt{R}
    & 45 & 384 & 10
    & 99 & 198 & 18
    & 100 & \best{85} & \best{9}
    \\
    \texttt{L}
    & 45 & \worst{1171} & \worst{13}
    & 99 & \worst{1080} & \worst{85}
    & 100 & \worst{664} & \worst{44}
    \\
    \texttt{S}
    & 43 & \worst{1627} & \worst{13}
    & 99 & \worst{1163} & \worst{124}
    & 100 & \worst{696} & \worst{45}
    \\
    \texttt{I}
    & 43 & 400 & 11
    & 99 & \best{175} & 19
    & 100 & \best{73} & 15
    \\
    \texttt{C}
    & 46 & 451 & 9
    & 99 & 226 & 21
    & 100 & 106 & \best{9}
    \\
    \texttt{U}
    & 45 & \best{368} & 10
    & 99 & 192 & \best{17}
    & 100 & 98 & 10
    \\
    \texttt{UI}
    & 42 & 394 & 10
    & 99 & \best{185} & 21
    & 100 & 94 & 12
    \\
    \texttt{US}
    & 44 & \best{358} & 10
    & 99 & \best{185} & 19
    & 100 & 94 & 11
    \\
    \texttt{UC}
    & 45 & 414 & 10
    & 99 & 208 & 18
    & 100 & 104 & 13
    \\
    \texttt{UR}
    & 44 & \best{370} & 10
    & 99 & 204 & \best{17}
    & 100 & \best{80} & \best{9}
    \\ \bottomrule
  \end{tabular}
\end{table}

The differences in the percentages in the first row are due to memory running out in some cases, but
they are small enough as to be statistically irrelevant.
In later rows, most networks are unextractable; the interesting observation here is that strategies
prioritising actions that involve two processes and unmarked processes tend to fail faster.

\paragraph{The unroller.}
Projections of choregraphies are intuitively easy to extract because their recursive procedures are
all synchronised (they come from the same choreography).
In practice, this is not necessarily the case: programs often include ``loops-and-a-half'', where it
is up to the programmer to decide where to place the duplicate code; and sometimes procedure
definitions can be locally optimised.
For example: if $X=\com peqx;\com p{e'}qx;\com p{e''}ry;X$, then in the extracted implementation of
$\pid q$ the definition of $X_{\pid q}$ can simply be $\brecv px;X$.

Our unroller models these situations by choosing one process and randomly unfolding some procedures,
as well as shifting the closing point of some loops.
These transformations are always correct, so they should yield extractable networks, but extraction
time may be larger and there may be higher chance for bad loops.
We generated $240$ tests, which we were all able to extract, and compared the extraction times for
the original and unrolled networks.
In Table~\ref{tab:unroller} we report the average and median ratios for each extraction strategy.

\begin{table}
  \caption{Extracting unrolled networks: for each strategy we report the average and the median of
    the ratio between the time needed to extract unrolled networks and the time needed to extract
    the original networks.}
  \label{tab:unroller}
  \centering
  \begin{tabular}{c|rr}
    \toprule
    Strategy
    & Average
    & Median
    \\ \midrule
    \texttt{R} & 6.20 & 1 \\
    \texttt{L} & 1.54 & 1 \\
    \texttt{S} & 4.82 & 1 \\
    \texttt{I} & 5.24 & 1.07 \\
    \texttt{C} & 1.95 & 1.06 \\
    \texttt{U} & 9.12 & 1.03 \\
    \texttt{UI} & 4.70 & 1 \\
    \texttt{US} & 2.17 & 1 \\
    \texttt{UC} & 3.88 & 1 \\
    \texttt{UR} & 2.47 & 1 \\
    \bottomrule
  \end{tabular}
\end{table}

The table shows that unrolling slows down the extracter somewhat, but in a very asymmetric way: for
most networks the changes are minor (shown by the median around $1$), while for a few there are very
large changes in either direction.
An analysis of the raw data shows that:
\begin{itemize}
\item there is no general trend -- in some cases the unrolled network is fastest to extract, in
  other cases it is slower;
\item in most cases the ratio is close to $1$ (and in many exactly $1$, due to the fact that
  execution times are rounded to the nearest millisecond);
\item ratios vary from as low as $0.0003$ to as high as $1505$.
\end{itemize}

\section{Conclusions and Discussion}
\label{sec:discussion}
\label{sec:concl}

We have presented an efficient algorithm for extracting choreographies from network specifications, improving the original conference presentation in~\cite{CLM17}.
We have successfully implemented this algorithm, developed the first comprehensive test suite for evaluating this kind of algorithms, and used the test suite to evaluate our implementation.
Our results are very encouraging compared to previous work~\cite{LTY15}, and open the door to interesting future developments.
We discuss some of them.

\paragraph{More expressive communications and processes.}
In real-world contexts, values stored and communicated by processes are typed, and the receiver
process can also specify how to treat incoming messages~\cite{CM17a}.
This means that communication actions now have the form $\com peqf$, where $f$ is the function
consuming the received message, and systems may deadlock because of typing errors.
Our construction applies without changes to this scenario -- any requirements regarding type
checking, for example, will also be necessary for defining the semantics of the process calculus.

\paragraph{Choreographic Programming and Multiparty Session Types.}
Choreographic languages like ours are used in choreographic programming, a programming paradigm where choreographies are programs that can be compiled to distributed implementations~\cite{M13:phd,CM20,M22}.
Our extraction algorithm can be applied to several existing languages for networks, modulo minor syntactic differences~\cite{CM17a,CM20,CMP18,M22}. For some of these languages, our algorithm can be applied only to fragments of them; we point out some of the features for future work in the next paragraphs.

Choreographies have also been advocated for the specification of communication
protocols.
Most notably, multiparty session types use choreographies to define types used in the verification of process calculi~\cite{HYC16}.
While there are multiple variants of multiparty session types, the one used most in practice so far is almost identical to a simplification of SP.
In this variant, each pair of participants has a dedicated channel, and communication actions refer
directly to the intended sender/recipient as in SP (see the theory
of~\cite{CM13,MY13,CLMSW16,CDYP16}, for example, and the practical implementations
in~\cite{HMBCY11,NY15,M13:phd}).
To obtain multiparty session types from SP (and CC), we just need to: remove the capability of
storing values at processes; replace message values with constants (representing types, which could
also be extended to subtyping in the straightforward way); and make conditionals nondeterministic
(since in types we abstract from the precise values and expression used by the evaluator).
These modifications do not require any significant change to our approach since our AES already
abstracts from data and, thus, our treatment of the conditional is already nondeterministic.
For reference, we can simply treat the standard construct for an internal choice at a process
$\pid p$ -- $\actor p{B_1\oplus B_2}$ -- as syntactic sugar for a local conditional such as
$\actor p{\bcond{\m{coinflip}}{B_1}{B_2}}$.

\paragraph{Asynchrony.}
Our process calculus is not expressive enough to model examples from~\cite{LTY15} that use the
pattern of asynchronous exchange.

\begin{example}
  \label{ex:async-mot}
  The network $\actor p{\bsend qe;\brecv qx} \parp \actor q{\bsend p{e'};\brecv py}$ is deadlocked
  in SP, but would run without errors in an asynchronous context: both $\pid p$ and $\pid q$ can
  send their respective values, becoming ready to receive each other's messages.
  \eoe
  \end{example}

Asynchronous semantics for SP and CC have been described in~\cite{CM17b}.
For SP, we add a FIFO queue for each pair of processes.
Communications now synchronise with these queues: send actions append a message in the queue of the
receiver, and receive actions remove the first message from the queue of the receiver.

In order to extract asynchronous exchanges, we do not need full asynchrony at the choreography
level. Rather, we can restrict ourselves to a new primitive called a \emph{multicom}~\cite{CMP18}: a list of communication actions with distinct receivers, written
$\genmulticometa$.
Using multicoms, the program in Example~\ref{ex:async-mot} can be extracted as
$\multicom{\com peqx, \com q{e'}py}$.
The theory of this extension has been discussed briefly in~\cite{CLM17}, but implementing it is outside of the scope of this work.

\paragraph{Process spawning.}
Another useful construct is the capability to spawn new processes at runtime~\cite{CHY12,CM17a}. This feature would suffice, for example, to represent the remaining examples from~\cite{LTY15}, as well as many more complex examples.
Having such a construct breaks the fundamental premise of our algorithm, namely that SEGs are finite. Studying how the theory and implementation could be adapted to this extension is a challenging future direction.

\paragraph{Extraction strategies.}
We also believe that extraction strategies have unexplored potential, but a full study of their impact goes beyond the scope of this work. An interesting direction could be to develop more complex heuristics, for example such that the choice of action to be consumed takes into account the shape of the network and the partial graph built so far.

%
%

\section*{Acknowledgments}
All authors were supported in part by the Independent Research Fund Denmark, Natural Sciences, grant DFF-7014-00041. Larsen was supported in part by the Independent Research Fund Denmark, Natural Sciences, grant DFF-0135-00018B. Montesi was supported in part by Villum Fonden, grant 29518, and the Independent Research Fund Denmark, Technology and Production, grant DFF-4005-00304.

\clearpage

\bibliographystyle{plain}
\bibliography{biblio}

\clearpage

\appendix

\end{document}